\documentclass[fleqn,usenatbib]{mnras}

\usepackage[T1]{fontenc}
\usepackage{ae,aecompl}

\usepackage{graphicx}	
\usepackage{amsmath}	
\usepackage{amssymb}	
\usepackage{xcolor}
\usepackage[normalem]{ulem}
\usepackage{booktabs}
\hypersetup{draft}
\definecolor{gray}{rgb}{0.0, 0.42, 0.24}
\definecolor{cobalt}{rgb}{0.0, 0.28, 0.67}
\definecolor{green}{rgb}{0.0, 0.55, 0.55}

\usepackage{threeparttable}
\usepackage{arydshln}

\title[Dust-gas chemistry in AGB outflows]{Chemical modelling of dust-gas chemistry within AGB outflows 
\\ III. Photoprocessing of the ice and return to the ISM}

\author[Van de Sande et al.]{
M. Van de Sande$^{1}$\thanks{E-mail: marie.vandesande@kuleuven.be}\thanks{Postdoctoral Fellow of the Fund for Scientific Research (FWO), Flanders, Belgium},
C. Walsh$^{2}$ \&
T.J. Millar$^{3}$
\\
$^{1}$Institute of Astronomy, KU Leuven, Celestijnenlaan 200D, 3001 Leuven, Belgium\\
$^{2}$School of Physics and Astronomy, University of Leeds, Leeds LS2 9JT, UK\\
$^{3}$Astrophysics Research Centre, School of Mathematics and Physics, Queen's University Belfast, University Road, Belfast BT7 1NN, UK
}

\date{Accepted XXX. Received YYY; in original form ZZZ}

\pubyear{2020}

\begin{document}
\label{firstpage}
\pagerange{\pageref{firstpage}--\pageref{lastpage}}
\maketitle

\begin{abstract}
To explain the properties of dust in the interstellar medium (ISM), the presence of a refractory organic mantle is necessary.
The outflows of AGB stars are among the main contributors of stellar dust to the ISM.
We present the first study of the refractory organic contribution of AGB stars to the ISM.
Based on laboratory experiments, we included a new reaction in our extended chemical kinetics model: the photoprocessing of volatile complex ices into inert refractory organic material.
The refractory organic feedback of AGB outflows to the ISM is estimated using observationally motivated parent species and grids of models of C-rich and O-rich outflows.
Refractory organic material is mainly inherited from the gas phase through accretion onto the dust and subsequent photoprocessing.
Grain-surface chemistry, initiated by photodissociation of ices, produces only a minor part and takes place in a sub-monolayer regime in almost all outflows.
The formation of refractory organic material increases with outflow density and depends on the initial gas-phase composition.
While O-rich dust is negligibly covered by refractory organics, C-rich dust has an average coverage of $3-9\%$, but can be as high as $8-22\%$.
Although C-rich dust does not enter the ISM bare, its average coverage is too low to influence its evolution in the ISM or significantly contribute to the coverage of interstellar dust.
This study opens up questions on the coverage of other dust-producing environments.
It highlights the need for an improved understanding of dust formation and for models specific to density structures within the outflow.
\end{abstract}




\section{Introduction}

Interstellar dust is thought to have a layered structure of a silicate core, an inner organic mantle, and an outer ice mantle.
The inner organic mantle consists of refractory organic material containing a mixture of aliphatic and aromatic carbonaceous molecules, formed by the photoprocessing of interstellar ices during their evolution in the interstellar medium (ISM) \citep{Greenberg1999,Jones2013}.
This inner refractory organic grain mantle is necessary to explain the amount and wavelength dependence of extinction, for which silicate grains alone do not suffice. 
A combination of silicates with graphite can mediate this \citep{Mathis1979}, but the polarisation of the extinction demands a coating of refractory organic material containing O, C and N. \citep{Greenberg1986}. 
Observational evidence comes mainly from the aliphatic CH stretching-mode band at 3.4 $\mu$m \cite[e.g.,][]{Willner1979,Butchart1986,Sandford1991,Duley1998,Pendleton2002}.
It has also been proposed as a potential carrier of the unidentified depleted oxygen in the ISM \citep{Whittet2010}.

Laboratory experiments have studied the formation of refractory organic material by exposing interstellar ice analogues to UV radiation.
When heating up the processed dust analogues, refractory organic material containing large molecules, carbonaceous material and even amino acids, stays behind \cite[e.g.,][]{Greenberg1972,Hagen1979,Allamandola1988,Bernstein2002,MunozCaro2003,Materese2017}.
The refractory organic mantle has been found to change the properties of the dust. 
For example, coated silicates have a much higher sticking threshold velocity \citep{Kouchi2002}, and the coating can increase the tensile strength of the dust grains while not inherently changing its stickiness \citep{Bischoff2020}.

The outflows of Asymptotic Giant Branch (AGB) stars are important contributors to the chemical enrichment of the ISM. 
Together with supernovae, they are the main contributors of pristine stellar dust to the ISM \citep{Zhukovska2013}.
The outflow is thought to be launched in a two-step mechanism, where stellar pulsations facilitate dust formation, which then launches a dust-driven wind \citep{Hofner2018}.
Exactly how the dust is formed is still unknown.
This implies that several characteristics of AGB dust are not known as it enters the ISM, such as its exact composition and its grain-size distribution.
These characteristics are crucial to our understanding of the physical and chemical dust evolution cycle in the ISM.
 
The chemical kinetics model developed in \citet{VandeSande2019b,VandeSande2020} (henceforth Papers I and II, respectively) is the first to consider dust-gas interactions and grain-surface chemistry in AGB outflows.
We found that gas-phase species can be efficiently depleted onto the dust, forming an ice mantle.
However, the chemical reaction network included only a simple surface chemistry, which limits the possible complexity of ice species.
The model also did not take the irradiation of the ice mantle into account and any photoprocessing that could occur in the outer regions of the outflow.
Given that our previous models predict efficient depletion of gas-phase species onto the dust, this raises the possibility of AGB outflows hosting a feedstock of volatile organic molecules on grain surfaces that could be photoprocessed into refractory organic material, as found in laboratory experiments.

Here, we present the first study of the contribution of refractory organic material from AGB outflows to the ISM.
To do so, a complex surface chemistry was included, along with photodissociation on the grain surface and reactive desorption.
We introduce photoprocessing of complex volatile ice species into inert refractory organic material, a first in chemical modelling of AGB outflows, using the results of laboratory experiments.
Based on observational samples of AGB stars, we assembled grids of C-rich and O-rich outflows.
We also compiled new observationally motivated sets of parent species.
This way, we are able to provide the first estimation of the refractory organic coverage of AGB dust as it enters the ISM, forming the starting point of its subsequent interstellar evolution.
This gives a more comprehensive view on the starting point of dust evolution in the ISM.

The new chemical reaction network and the new reactions included in the model are described in Sect. \ref{sect:model}, along with the compilation of observational grids of C-rich and O-rich outflows and of the observationally motivated sets of parent species.
Our results on the formation of refractory organic material within AGB outflows and their contribution to the ISM are shown in Sect. \ref{sect:results}.
This is followed by their discussion in Sect. \ref{sect:discussion} and the conclusions in Sect. \ref{sect:conclusions}.

\section{The chemical model}				\label{sect:model}

\begin{figure*}
 \includegraphics[width=0.9\textwidth]{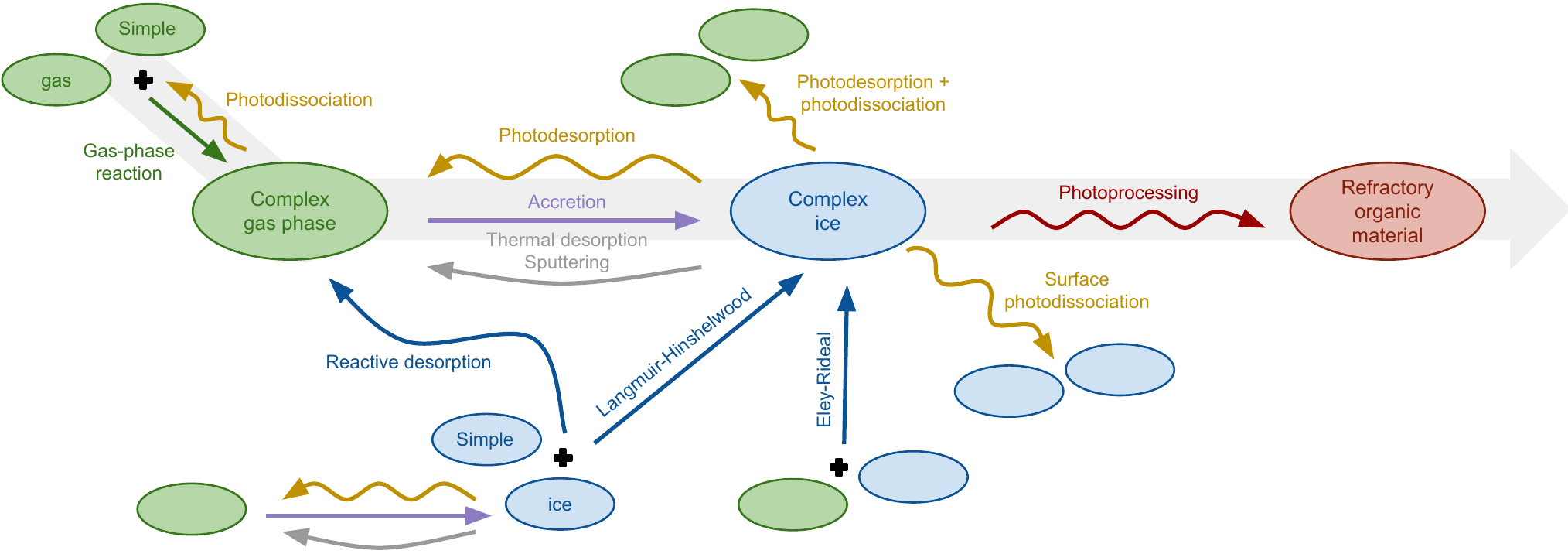}
 \caption{Scheme representing the different formation and destruction pathways included in the chemical network.
 Gas-phase reactions are simplified in this scheme.
 Photoprocessing of complex ice species into inert refractory organic material creates a sink of C-bearing material on the dust grain.
  Green ellipses: gas-phase species, blue ellipses: ice species, red ellipse: refractory organic material. 
  The main pathway to the formation complex ices and refractory organic material is highlighted by the grey arrow.
   }
 \label{fig:scheme}
\end{figure*}

The one-dimensional chemical kinetics model is fully described in Papers I and II.
Details of the model can be found in Appendix \ref{app:model}.
We use the chemical reaction network of \citet{Walsh2014} designed for use in protoplanetary disk chemical models.
This includes the full gas-phase network UDfA \textsc{Rate12} \citep{McElroy2013} supplemented with a set of grain-surface reactions and associated processes, including needed gas-phase reactions, extracted from the 2008 version of Ohio State University (OSU) network \citep{Garrod2008}.
The surface-chemistry network includes atom-addition reactions to form both simple (e.g., water, H$_2$O)  and more complex ices (e.g., methanol, CH$_3$OH) as well as radical-radical recombination reactions that help to build larger molecules (e.g., glycolaldehyde, (CH$_2$OH)$_2$).
The network also includes thermal and non-thermal desorption processes (photodesorption, reactive desorption, and sputtering) and includes the dissociation of ice species on the surface via photodissociation and photoionisation by both external photons and those generated internally by cosmic rays.

Table \ref{table:modelparams} lists all parameters describing the outflow and dust.
The gas temperature profile does not change throughout the modelling.
The gas-to-dust mass ratio and the surface density of binding sites are kept constant.
The dust-grain size distribution (GSD) is fixed to the canonical MRN distribution \citep{Mathis1977}.
The dust bulk density depends on whether the outflow is C-rich or O-rich.
The outer radius of the model depends on the outflow density, and is taken to be where the H$_2$ number density reaches 10 cm$^{-3}$, i.e., where the outflow is assumed to merge with the ISM.

The new grain-surface reactions included in the model are described in Sect. \ref{subsect:model:reactions}.
The observational set of AGB outflows is described in Sect. \ref{subsect:model:obs}, together with the observational motivation behind the assumed drift velocities and dust compositions.
The new observationally motivated sets of parent species are presented in Sect. \ref{subsect:model:parents}.

\begin{table}
	\caption{Parameters of the grid of chemical models.}
	\centering
	\label{table:modelparams}
	\begin{tabular}{ll} 
		\hline
	\multicolumn{2}{c}{Fixed parameters} \\
	\hline
    \multicolumn{2}{c}{Physical parameters} \\
    \cdashline{1-2}
    \noalign{\smallskip}
    Stellar temperature, $T_*$        & 2000 K \\
    Stellar radius, $R_*$             & 5 $\times 10^{13}$ cm \\
    Exponent $T_\mathrm{gas}(r)$, $\epsilon$                       & 0.7 \\
    Initial radius of the model		& $10^{15}$ cm \\
    Final radius of the model		& $10^{18}$ cm \\
    \noalign{\smallskip}
    \multicolumn{2}{c}{Dust parameters} \\
    \cdashline{1-2}
    \noalign{\smallskip}
	Minimum grain size, $a_\mathrm{min}$ & $5 \times 10^{-7}$ cm \\
	Maximum grain size, $a_\mathrm{max}$ & $2.5 \times 10^{-5}$ cm \\
	Dust-to-gas mass ratio, $\psi$			& $2 \times 10^{-3}$ \\
	Surface density of binding sites, $n_s$	& $10^{15}$ cm$^{-2}$ \\
	Silicate dust bulk density$^1$	& 3.5 g cm$^{-3}$  \\
	Carbonaceous dust bulk density$^1$ 	& 2.24 g cm$^{-3}$  \\
    \hline
    	\end{tabular}
    	\centering
	\resizebox{1.0\columnwidth}{!}{%
	\begin{tabular}{ll} 
    	\multicolumn{2}{c}{Variable parameters} \\
	\hline
    \noalign{\smallskip}
    \multicolumn{2}{c}{Physical parameters} \\
    \cdashline{1-2}
    \noalign{\smallskip}
    Mass-loss rate, $\dot{M}$        &    \\
    Outflow velocity, $v_\infty$   &   \\
    Drift velocity, $v_\mathrm{drift}$   &  5 km s$^{-1}$ ($\dot{M} \geq 10^{-5}$ M$_\odot$ yr$^{-1}$) \\
                                   &  10 km s$^{-1}$ ($10^{-7} < \dot{M} < 10^{-5}$ M$_\odot$ yr$^{-1}$) \\
                                   &  15 km s$^{-1}$ ($\dot{M} \leq 10^{-7}$ M$_\odot$ yr$^{-1}$) \\
    \noalign{\smallskip}
    \multicolumn{2}{c}{Dust parameters} \\
    \cdashline{1-2}
    \noalign{\smallskip}
	Prefactor $T_\mathrm{dust}(r)$, $T_\mathrm{dust,*}$ & \\
	Exponent $T_\mathrm{dust}(r)$, $s$ & \\
	\hline
	\end{tabular}
	}
	\footnotesize
    { {{References.}} (1) \citet{Draine2003}
    }
\end{table}

\subsection{New grain-surface reactions}			\label{subsect:model:reactions}

Including a more complex chemistry requires adding grain-surface reactions to render the network more complete: reactive desorption and photodissociation of ices on the surface.
This is done following the recipes of \citet{Cuppen2017} for models using the mean-field rate-equation approach.
The influence of these new reactions on the chemistry is discussed in Appendix \ref{app:influencenew}.
We also include a new type of reaction, the photoprocessing of volatile ice species to inert refractory organic material.
Fig. \ref{fig:scheme} illustrates all pathways included in the network, focussing on the grain-surface chemistry.

\paragraph*{Reactive desorption}        

Grain-surface reactions, via either the diffusive Langmuir-Hinshelwood or the stick-and-hit Eley-Rideal mechanism, can result in desorption of the product into the gas phase because of excess reaction energy.
The reactive desorption rate is calculated as
\begin{equation}
	k_\mathrm{RD} = B\ k_\mathrm{LH/ER}\ \ \mathrm{s^{-1}},
\end{equation}
where $k_\mathrm{LH/ER}$ is the Langmuir-Hinshelwood or the Eley-Rideal reation rate (Paper I) and $B$ is the reactive desorption efficiency.
We assume that $B = 0.01$ throughout our calculations.

\paragraph*{Photodissociation on the grain surface}

Photodissociation on the grain surface is treated as a parallel process to photodesorption.
The reaction rates of photodissociation caused by UV photons or cosmic-ray induces photoreactions are those given in \citet{McElroy2013}.

\begin{figure}
 \includegraphics[width=1.\columnwidth]{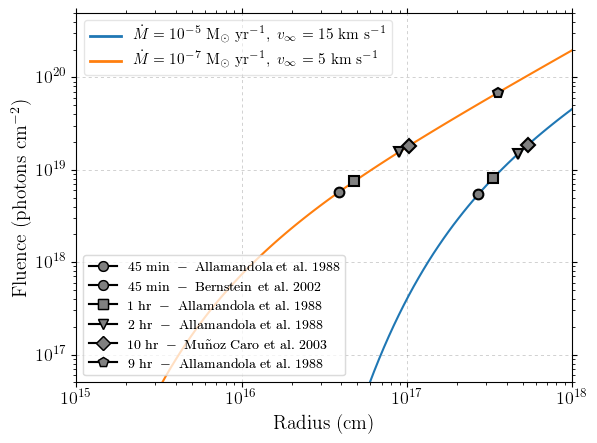}
 \caption{Fluence throughout a high density outflow ($\dot{M} = 10^{-5}$ M$_\odot$ yr$^{-1}$, $v_\infty = 15$ km s$^{-1}$, blue line) and a low density outflow ($\dot{M} = 10^{-7}$ M$_\odot$ yr$^{-1}$, $v_\infty = 5$ km s$^{-1}$, orange line).
  Markers: fluences obtained in lab experiments after a certain exposure time.
  }
 \label{fig:fluence}
\end{figure}

\paragraph*{Photoprocessing to refractory organic material}

The photoprocessing of volatile ice to inert refractory organic material is limited to complex species only, which are defined as species containing at least three C atoms or at least two C atoms and one other heavy atom (N, O, Si, or S).
No back reaction is included for refractory organic material, so that photoprocessing creates a sink for complex ice species and removes any further possible reactions between them.
Polymerisation is not explicitly included in the network, but is implied as part of the conversion from volatile to refractory organics.
The refractory organics do not react among themselves, nor with ice and gas-phase species, and hence are being treated as chemically inert.
The photoprocessing rate is similar to that of photodissociation,
\begin{equation}
	k_\mathrm{VR} = \omega\ \exp\left( -\gamma A_V\right)\ \ \mathrm{s^{-1}},
\end{equation}
where $\omega$ is the photoprossessing rate to refractory organic material in the unshielded interstellar UV field, $A_V$ is the dust extinction at visible wavelengths, and $\gamma$ is a parameter that takes the increased dust extinction at UV wavelengths into account \citep{McElroy2013}.
We have assumed that $\gamma$ is equal to that of H$_2$O, which is generally the largest ice component. 
This value is also similar to many C-bearing molecules in ices.

The value of $\omega$ is empirically estimated from laboratory experiments.
Fig. \ref{fig:fluence} shows fluences used in experiments by \citet{Allamandola1988}, \citet{Bernstein2002}, and \citet{MunozCaro2003} together with the fluence of interstellar UV photons in a low- and high-density outflow.
This is calculated by integrating the flux of interstellar UV photons throughout the outflow over time, done by multiplying the flux at a specific radius within the outflow by the expansion time, which is the radius divided by the outflow velocity.
The fluences needed to convert volatile ices to inert refractory organic material are reached in the outer region, at $\sim 10^{17}$ cm, for both outflow densities.
The fluence corresponding to a 37 h exposure \citep{Allamandola1988} is achieved outside the outflow and not visible in Fig. \ref{fig:fluence}.
From the exposure times, we find that the photoprocessing rate $\omega$ lies between $3 \times 10^{-4} - 8 \times 10^{-6}$ s$^{-1}$.
We have taken $\omega = 10^{-5}$ s$^{-1}$ as a reasonable value for the photoprocessing rate.

\begin{table}
	\caption{Parent species and their initial abundances relative to H$_2$. Upper part: O- and C-rich parents of Papers I and II. 
	Lower part: C-rich parent species as derived from observations \citep{Agundez2020}.
	Left column: mean abundances (ab.) from the observed ranges. Right column: mean abundances, with the maximal observed abundance for C-bearing species.
} 
 	\centering
    \begin{tabular}{l c c l c }
    \hline  
    \multicolumn{5}{c}{Parent species used in Papers I \& II} \\
	\hline
    \noalign{\smallskip}
    \multicolumn{2}{c}{Oxygen-rich} & & \multicolumn{2}{c}{Carbon-rich}  \\
    \cline{1-2} \cline{4-5} 
    \noalign{\smallskip}
 He        &	0.17						&	&	    He            	&		0.17						\\
 CO        &	 $3.0\times10^{-4}$    	&	&	    CO            	&	 $8.0\times10^{-4}$   	\\
 N$_2$     &	 $4.0\times10^{-5}$    	&	&	    N$_2$         	&	 $4.0\times10^{-5}$   \\
 H$_2$O    &	 $3.0\times10^{-4}$    	&	&	    C$_2$H$_2$    	&	 $8.0\times10^{-5}$    	\\
 CO$_2$    &	 $3.0\times10^{-7}$    	&	&	    HCN           	&	 $2.0\times10^{-5}$    	\\
 SiO       &	 $5.0\times10^{-5}$  	&	&	    SiO           	&	 $1.2\times10^{-7}$   \\
 SiS       &	 $2.7\times10^{-7}$  	&	&	    SiS           	&	 $1.0\times10^{-6}$    	\\
 SO        &	 $1.0\times10^{-6}$    	&	&	    CS            	&	 $5.0\times10^{-7}$  	\\
 H$_2$S    &	 $7.0\times10^{-8}$    	&	&	    SiC$_2$       	&	 $5.0\times10^{-8}$    	\\
 PO        &	 $9.0\times10^{-8}$    	&	&	    HCP           	&	 $2.5\times10^{-8}$   \\
 HCN       &	 $2.0\times10^{-7}$  	&	&	    NH$_3$        	&	 $2.0\times10^{-6}$  	\\
 NH$_3$    &	 $1.0\times10^{-7}$  	&	&	    H$_2$O        	&	 $1.0\times10^{-7}$    	\\
    \hline 
    \end{tabular}%
 	\centering
    \begin{tabular}{l c cc}
    \hline  
    \multicolumn{4}{c}{Parent species compiled from \citet{Agundez2020}}\\
    \hline 
    \noalign{\smallskip}
     & Carbon-rich, & & Carbon-rich,\\
     & mean observed ab.& & mean observed ab.,\\
     &  & & max. C-bearing \\
    \cline{1-2} \cline{4-4} 
    \noalign{\smallskip}
He			&	$1.70\times 10^{-1}$	& &	$1.70\times 10^{-1}$	\\
CO			&	$8.00\times 10^{-4}$	& &	$8.00\times 10^{-4}$	\\
N$_2$		&	$4.00\times 10^{-5}$	& &	$4.00\times 10^{-5}$	\\
C$_2$H$_2$	&	$4.38\times 10^{-5}$	& &	$8.00\times 10^{-5}$	\\
HCN			&	$4.09\times 10^{-5}$	& &	$8.00\times 10^{-5}$	\\
SiO			&	$5.02\times 10^{-6}$	& &	$5.02\times 10^{-6}$\\
SiS			&	$5.98\times 10^{-6}$& &	$5.98\times 10^{-6}$	\\
CS			&	$1.06\times 10^{-5}$	& &	$2.10\times 10^{-5}$	\\
SiC$_2$		&	$1.87\times 10^{-5}$	& &	$3.70\times 10^{-5}$	\\
HCP			&	$2.50\times 10^{-8}$	& &	$2.50\times 10^{-8}$\\
NH$_3$		&	$6.00\times 10^{-8}$	& &	$6.00\times 10^{-8}$	\\
H$_2$O			&	$2.55\times 10^{-6}$	& &	$2.55\times 10^{-6}$\\
CH$_4$		&	$3.50\times 10^{-6}$	& &	$3.50\times 10^{-6}$\\
HCl			&	$3.25\times 10^{-7}$	& &	$3.25\times 10^{-7}$	\\
C$_2$H$_4$	&	$6.85\times 10^{-8}$	& &	$1.00\times 10^{-7}$	\\
HF			&	$1.70\times 10^{-8}$	& &	$1.70\times 10^{-8}$	\\
H$_2$S			&	$4.00\times 10^{-9}$	& &	$4.00\times 10^{-9}$	\\
    \hline 
    \end{tabular}%
    \label{table:parentspecies}    
\end{table}

\begin{figure}
 \includegraphics[width=1.\columnwidth]{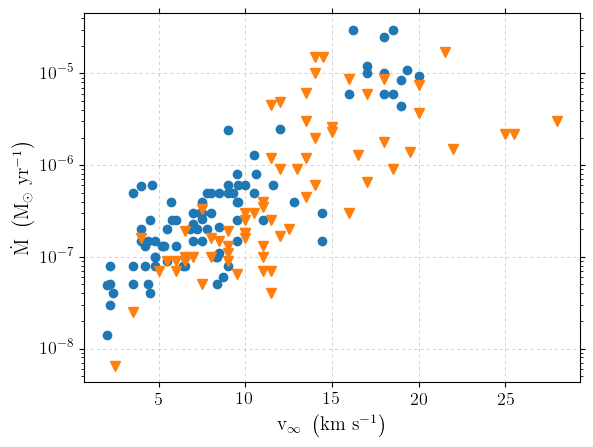}
 \caption{Measured mass-loss rate, $\dot{M}$, and outflow velocity, $v_\infty$, of the AGB outflows included in the observational grids. 
 Blue circles: O-rich outflows \citep{Olofsson2002,GonzalezDelgado2003,Danilovich2015}.
 Orange triangles: C-rich outflows \citep{Schoier2001,Danilovich2015}.
  }
 \label{fig:observations}
\end{figure}

\subsection{Observational grids of AGB outflows}			\label{subsect:model:obs}

In Papers I and II, we assumed representative outflow densities for the model calculations.
In this work, we used observational studies of AGB outflows, where mass-loss rate and outflow velocity were retrieved, to compile grids of O-rich and C-rich outflows.
These observationally motivated grids are necessary to estimate the refractory organic output of AGB stars to the ISM.

The sample of O-rich outflows is taken from \citet{Olofsson2002}, \citet{GonzalezDelgado2003}, and \citet{Danilovich2015}.
For C-rich outflows, we used \citet{Schoier2001} and \citet{Danilovich2015}.
If a star was present in multiple datasets, we have taken the measured mass-loss rate and outflow velocity from the most recent one.
Fig. \ref{fig:observations} shows the compiled O-rich and C-rich observational data sets.

The drift velocity between dust and gas is not readily retrieved from observations.
Therefore, only a few outflows have a measured drift velocity \cite[e.g.,][]{Groenewegen1997,Decin2010,VandeSande2018a}.
In general, higher mass-loss rates are expected to have lower drift velocities.
Together with the sample of \citet{Ramstedt2008}, we have chosen for $v_\mathrm{drift} = 5$ km s$^{-1}$ if $\dot{M} \geq 10^{-5}$ M$_\odot$ yr$^{-1}$, $v_\mathrm{drift} = 10$ km s$^{-1}$ if $10^{-7} < \dot{M} < 10^{-5}$ M$_\odot$ yr$^{-1}$, and $v_\mathrm{drift} = 15$ km s$^{-1}$ if $\dot{M} \leq 10^{-7}$ M$_\odot$ yr$^{-1}$.

The assumed dust composition is also based on observations.
\citet{Heras2005} derived the composition of the dust around 28 O-rich AGB stars and found that Fe-bearing silicates are present in all outflows.
Its total fraction depends on the outflow, as does the general composition of the dust.
For the O-rich models, we assume that the dust is composed of a 50/50 mixture of melilite and silicate with Fe.
This simplification roughly accounts for some of the variation seen in O-rich dust. 
However, it is necessary to calculate models for the observational sets, as the dust composition is not known for each outflow.
For the C-rich models, we assume amorphous carbon as the only dust species.
While SiC dust has been observed around C-rich AGB stars \cite[e.g.,][]{Yang2004}, we can assume that its temperature is similar to that of amorphous carbon as their overall wavelength-dependence of absorption is similar \citep{Rau2017}.
The calculation of the dust temperature profiles is detailed in Appendix \ref{app:tdustparams}.

The measured mass-loss rates and outflow velocities, assumed drift velocities, and parameters of the dust temperature profiles are listed in Tables \ref{table:paramscrich} and \ref{table:paramsorich} for the grid of C-rich and O-rich outflows, respectively.

\subsection{Parent species}			\label{subsect:model:parents}

Table \ref{table:parentspecies} lists the sets of parent species used.
In addition to the O- and C-rich parent species of Papers I and II, we compiled two observationally motivated sets for C-rich outflows.
These are based on \citet{Agundez2020}, who have compiled (ranges of) observed abundances of parent molecules in C-rich, O-rich, and S-type stars (see their Table 2).
One set uses the mean of the ranges in observed abundances, as an approximation of an average C-rich outflow.
The other has the maximal abundances for C-bearing species to maximise the possible refractory organic feedstock, to explore an extreme case.

\begin{figure}
 \includegraphics[width=1.\columnwidth]{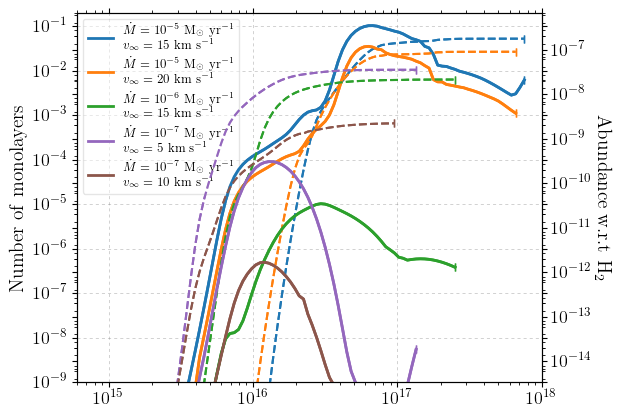}
 \caption{Number of monolayers (left y-axis) and abundance relative to H$_2$ (right y-axis) of volatile ice (solid lines) and refractory organic material (dashed lines) in C-rich outflows with different mass-loss rates and outflow velocities, using the parent species of Papers I and II.
 The dash marks where the H$_2$ number density reaches 10 cm$^{-3}$, i.e., where the outflow merges with the ISM.
  }
 \label{fig:outputcrich}
\end{figure}

\section{Results}				\label{sect:results}

Refractory organic material can be formed on the dust within AGB outflows.
The majority of refractory organic material is inherited from the gas phase: complex species are formed in the gas phase, accreted onto the dust and then photoprocessed into refractory organic material over the timescale of the outflow (highlighted in Fig. \ref{fig:scheme} by a grey arrow).
Complex ice species can also be formed through grain-surface chemistry.
They too can be subsequently photoprocessed, but make up only a minority of the refractory organic material injected into the ISM.

The outflow density and the initial gas-phase composition play a crucial role:
(i) The outflow density determines the accretion rate and the onset of photon-reactions (photodissociation in the gas phase and the ice, photodesorption and photoprocessing).
The onset shifts closer to the star as the outflow density decreases.
Before the onset of photon-reactions, the ice mantle is directly inherited from the gas phase through accretion. 
After the onset, the accreted complex ice species can be photoprocessed into refractory organic material.
They can also be photodissociated into simpler and radical ices, starting the minor grain-surface chemistry formation route to complex ices and subsequently refractory organic material.
(ii) The initial gas-phase composition determines the availability of gas-phase and icy feedstock material.
C-rich outflows inherently provide more feedstock.
The initial abundances of HCN and C$_2$H$_2$, especially, are essential.
Conversely, O-rich outflows show a much smaller refractory organic coverage, linked to dearth of C-bearing parent species.

The refractory organic output to the ISM is limited. 
O-rich dust is barely covered, while  $3-9\%$ of the surface of C-rich dust is on average covered by refractory organic material.
Nonetheless, photoprocessing of volatile ice to inert refractory organic material is essential to the long-term survival of the gas-phase species accreted and processed throughout the outflow.
Without it, the entire ice mantle would be ablated as it enters the ISM.

In Sects. \ref{subsect:results:crich} and \ref{subsect:results:orich}, we present the results for C-rich and O-rich outflows, respectively.
First, the formation of refractory organic material in representative outflows is shown, followed by a quantification of the refractory organic output of the observational grids to the ISM.

\subsection{C-rich outflows}			\label{subsect:results:crich}

\begin{table}
	\caption{Outflow-density weighted average refractory organic output of C-rich outflows for different parent species, together with the average number of refractory organic molecules per grain and their fractional abundance with respect to H$_2$ in the ISM.
} 
 	\centering
	\resizebox{1.0\columnwidth}{!}{%
    \begin{tabular}{l l l l}
    \hline  
	 & Papers I \& II & Mean obs. & Mean obs., \\
	 & & & max C \\
    \hline
    Mean N$_\mathrm{mono}$ &	$5.66\times 10^{-2}$ & $3.56\times 10^{-2}$	&  $9.45\times 10^{-2}$ \\
	Mean \# per grain &	$7.64\times 10^2$		& $4.80\times 10^2$		& $1.27\times 10^3$ \\
	Frac. ab. in ISM &	$7.64\times 10^{-10}$	&  $4.80\times 10^{-10}$	& $1.27\times 10^{-9}$\\
    \hline 
    \end{tabular}%
    }
    \label{table:outputobscrich}    
\end{table}

\begin{figure*}
 \includegraphics[width=0.95\textwidth]{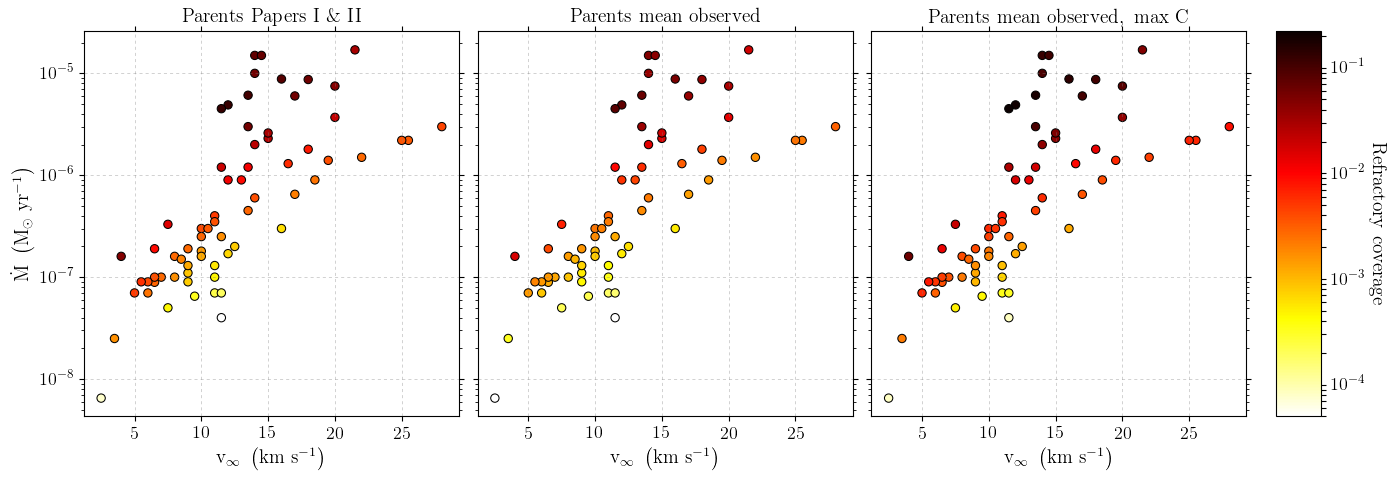}
 \caption{Refractory organic coverage of the dust grains in the outflows of the C-rich observational grid, for different sets of parent species (Table \ref{table:parentspecies}). 
 The coverage corresponds to the number of refractory organic monolayers where the outflow merges with the ISM, which is taken to be the radius where $n(\mathrm{H_2}) \leq 10$ cm$^{-3}$.
  Left panel: parent species used in Papers I and II. Middle panel: mean observed abundances from \citet{Agundez2020}. Right panel: mean observed abundances, with the maximal observed value for C-bearing species.
   }
 \label{fig:outputobscrich}
\end{figure*}

\begin{figure*}
 \includegraphics[width=1.0\textwidth]{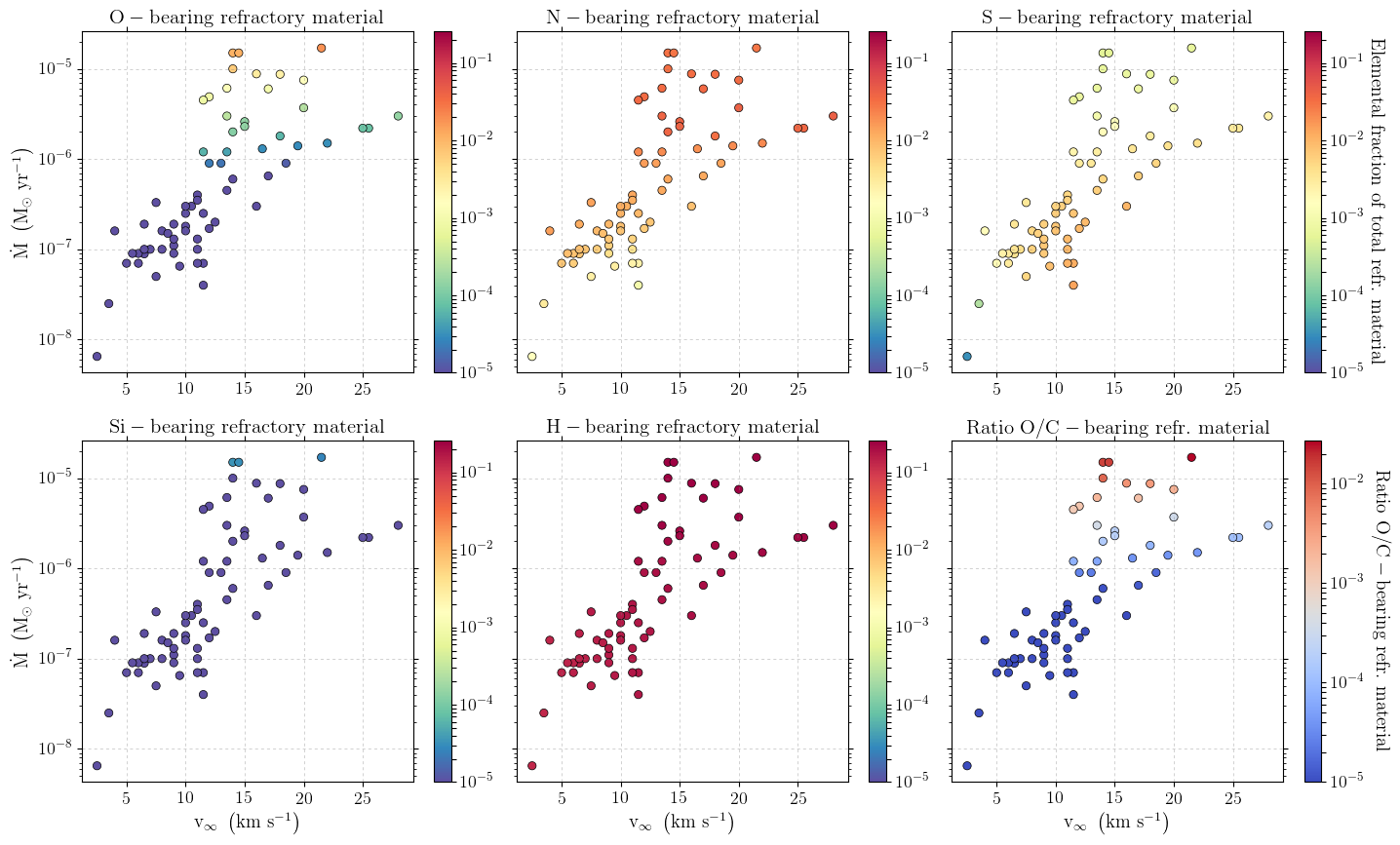}
 \caption{First five panels: elemental fractional composition of the total refractory organic output (O-, N-, S-, Si-, and H-bearing refractory organic material) for the outflows of the C-rich observational dataset.
 The parent species used are the mean observed abundances from \citet{Agundez2020}. 
 Lower right panel: ratio of O-bearing/C-bearing refractory organic material of the total refractory organic output.
   }
 \label{fig:outputobscrich-allfracs}
\end{figure*}

Fig. \ref{fig:outputcrich} shows the number of monolayers of volatile ice and inert refractory organic material in different C-rich outflows together with their total abundance relative to H$_2$, using the parent species of Papers I and II.
The outflow densities are based on the observational grid and are representative for C-rich AGB stars.
The influence of the outflow density can be clearly seen.
The highest density outflow has a peak number of ice monolayers of $\sim 0.1$, more than five orders of magnitude larger than in the lowest density outflow.
Surface chemistry hence takes place in a sub-monolayer regime.
Consequently, the highest density outflow has $\sim 0.05$ monolayers of refractory organic material, which corresponds to a coverage of $\sim 5\%$, while the lowest density outflow has a refractory organic coverage of less than $0.1 \%$.
The onset of photoprocessing shifts closer to the star with decreasing outflow density.
Feedstock ices are therefore photoprocessed more efficiently in lower density outflows, causing their rapid increase in refractory organic material.
Therefore, the number of refractory organic monolayers differs only by about two orders of magnitude between the highest and the lowest density outflows, compared to more than five orders of magnitude for the number of ice monolayers.

Fig. \ref{fig:outputobscrich} shows the refractory organic coverage of the dust in the C-rich observational grid, using the three sets of parent species (Table \ref{table:parentspecies}).
The coverage corresponds to the percentage of the surface covered by refractory organic material where the outflow merges with the ISM, which is taken to be the radius where $n(\mathrm{H_2}) \leq 10$ cm$^{-3}$.
The coverage increases with increasing outflow density, achieved via a larger mass-loss rate or a lower outflow velocity.
The maximum refractory organic coverage ranges from $8-22\%$, where the range is due to the different sets of parent species. 
We find that $25-35\%$ of the outflows in the C-rich grid have a refractory organic coverage larger than 1\%.
A coverage larger than 10\% is only achieved in up to 14\% of the outflows.
For CW Leo (IRC+10216), the most studied C-rich AGB star, we find a refractory organic coverage between $5-11\%$.

The dependency on the parent species is due to the relative abundances of C$_2$H$_2$ and HCN.
These are the main contributors of carbon and are crucial to the origin the gas-phase complexity \cite[e.g.,][]{Millar1994,Agundez2017}.
C$_2$H$_2$ is the key driver of chemical complexity, while HCN is necessary for the formation of N-bearing C-chains, e.g., cyanopolyynes.
The mean observed parents with maximal carbon yield the largest refractory organic coverage as they have the largest C$_2$H$_2$ and HCN abundances.
The parents of Papers I and II have the same C$_2$H$_2$ abundance but a lower HCN abundance, resulting in a smaller refractory organic coverage.
The mean observed parents have the lowest C$_2$H$_2$ abundance and therefore the smallest refractory organic coverage, even though the HCN abundance is larger than that in Papers I and II.

\begin{figure}
 \includegraphics[width=1.\columnwidth]{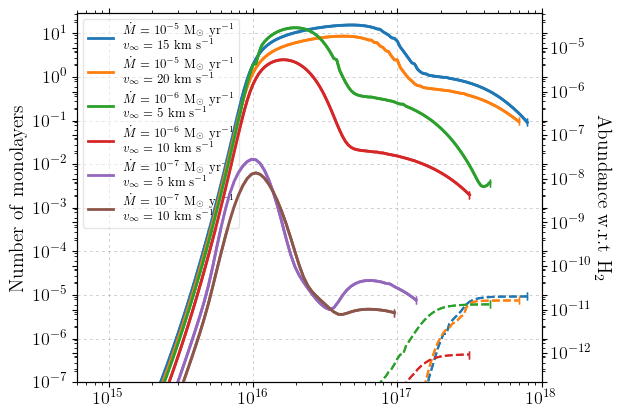}
 \caption{Number of monolayers (left y-axis) and abundance relative to H$_2$ (right y-axis) of volatile ice (solid lines) and refractory organic material (dashed lines) in O-rich outflows with different mass-loss rates and outflow velocities. 
 The dash marks where the H$_2$ number density reaches 10 cm$^{-3}$, i.e., where the outflow merges with the ISM.
  }
 \label{fig:outputorich}
\end{figure}

Table \ref{table:outputobscrich} summarises the results of Fig. \ref{fig:outputobscrich} and lists the outflow-density weighted average refractory organic output of the C-rich grid to the ISM.
The average is calculated by weighing the mean by the mass-loss rate divided by the outflow velocity.
The average number of refractory organic species per dust grain is calculated by multiplying the outflow-density weighted average number of monolayers by the average number of sites per grain.
In the MRN GSD, the average cross-section of a grain is $\langle \sigma_g \rangle = 3.37 \times 10^{-12}$ cm$^2$, so that the average number of sites per grain is equal to $\Sigma_\mathrm{sites} = 4 \langle \sigma_g \rangle n_s = 1.35 \times 10^4$.
Using the typical average dust density in the ISM  of $10^{-12}$ relative to H$_2$ \cite[e.g.,][]{Tielens2005}, this can be converted to a fractional abundance of refractory organic material originating from AGB outflows in the ISM.
The average outputs of the different sets of parent species lie within a factor two of each other.

Fig. \ref{fig:outputobscrich-allfracs} shows the elemental fractional composition of the refractory organic material in the C-rich outflows and its O/C ratio, using the mean observed set of parent species.
The elemental fractional composition is calculated per atom: for example, for H, it is the total number of H atoms over all species divided by the total number of all atoms over all species.
The amount of O-, N-, Si-, and H-bearing material increases with outflow density, while the amount of S-bearing material decreases.
The refractory organic layer is poor in Si, with generally less than 0.01\% of the material locked up in Si.
An O-bearing component of $\sim 2\%$ is only reached for the highest density outflows, with most outflows having a coverage lower than $\sim 0.01\%$. 
The N-bearing component ranges from $\sim 5\%$ to $\sim 0.1\%$.
The material is rich in H, with between $10 - 25\%$ of refractory organic material locked up in H.
The S-bearing fraction ranges from $\sim 1\%$ for the lower density outflows to $\sim 0.1\%$ for the higher density outflows.
The O/C ratio of the refractory organic material is generally low, reflecting the small fraction of O-bearing material. 
The ratio reaches values of approximately $10^{-2}$ only for the highest density outflows.

\begin{table}
	\caption{Outflow-density weighted average refractory organic output of O-rich outflows, together with the average number of refractory organic molecules per grain and their fractional abundance with respect to H$_2$ in the ISM.
} 
 	\centering
    \begin{tabular}{l l }
    \hline  
	 & Papers I \& II \\
    \hline
    Mean N$_\mathrm{mono}$ &	$6.28\times 10^{-6}$		 \\
	Mean \# per grain 	  &	$8.47\times 10^{-2}$		 \\
	Abundance in ISM 	  &	$8.47\times 10^{-14}$	\\
    \hline 
    \end{tabular}%
    \label{table:outputobsorich}    
\end{table}

\subsection{O-rich outflows}			\label{subsect:results:orich}

Fig. \ref{fig:outputorich} shows the number of monolayers of volatile ice and inert refractory organic material, together with their total abundance relative to H$_2$, for different O-rich outflows representative of the observational grid of O-rich AGB stars.
The peak number of ice monolayers is larger than for the C-rich outflow, ranging from 10 monolayers for the highest density outflows to $\sim10^{-2}$ monolayers  for the lowest.
This is due to the large initial H$_2$O abundance, which has a large binding energy of 4880 K (see Paper I).
The decrease in the ice mantle at large radii is caused by photodissociation on the grain surface.
It does not correspond to an increase in refractory organic material, as in C-rich outflows (Fig. \ref{fig:outputcrich}), because of the low carbon-content of the ice mantle.
Although the ices used in laboratory experiments on refractory organic formation are also H$_2$O-rich, their carbon content is typically greater than 1\%.
Only higher density outflows reach a refractory organic coverage of around $0.001\%$. 
Both in the parent species of Papers I and II and those listed in \citet{Agundez2020}, C$_2$H$_2$ is not present and the HCN abundance is at least two orders of magnitude smaller than in C-rich outflows.
Therefore only few C-bearing ices are present.

Table \ref{table:outputobsorich} lists the outflow-density weighted average refractory organic output of O-rich outflows to the ISM, summarising the results of the observational grid.
The average coverage is four orders of magnitude smaller than that of C-rich outflows. 
With on average 8.5 refractory organic species per dust grain, the coverage is negligible in the rate equation approach.
The largest coverage is only 0.0015\%.
The refractory organic coverage of the observational sample and its fractional composition are shown in Appendix \ref{app:outputorich}.

\section{Discussion}				\label{sect:discussion}

Including photoprocessing of complex ice species in the chemical model leads to the build-up of refractory organic material.
In AGB outflows, chemical complexity on the dust originates mainly from the gas phase: complex gas-phase species are accreted onto the dust and are subsequently photoprocessed.
Active grain-surface chemistry plays only a minor role in the formation of complex ice species and refractory organic material.
Complexity is not the result of simple ices reacting on the grain surface.
This is unlike other environments where gas-grain chemistry takes place, such as dark clouds and protoplanetary disks \cite[e.g.,][]{Henning2013,Tielens2013}.

The grain-surface chemistry (photodissociation of ices and surface reactions between them) occurs in a sub-monolayer regime for the majority of the outflows.
Sub-monolayer regimes are not common in astrochemical environments, holding only in the transition region from diffuse to dense interstellar clouds, where ice mantles are formed, and during the desorption of ice mantles at snowlines in protoplanetary disks \citep{Potapov2019}.

In Sect. \ref{subsect:discussion:complexity}, we discuss the origin of chemical complexity on the dust, together with the fractional composition of the refractory organic material.
Sect. \ref{subsect:discussion:output} focusses on the refractory organic contribution of AGB stars to the ISM and its impact on other astrochemical environments.
Sect. \ref{subsect:discussion:surface} discusses the possible influence of surface effects on the refractory organic output.
The sensitivity of our output to assumptions made in the model are discussed in Sect. \ref{subsect:discussion:uncert}.
The influence of the new chemical network on the gas-phase chemistry is discussed in Appendix \ref{app:gas}.

\subsection{Refractory organic composition in C-rich outflows}			\label{subsect:discussion:complexity}

The effects of the two formation processes, accretion and surface chemistry, are visible in Fig. \ref{fig:outputobscrich-allfracs}, which shows the fractional composition of the refractory organic material in the C-rich observational grid, using the mean observed set of parent species.
The same processes are also at play in the O-rich outflows. 
As these outflows have significantly lower refractory organic coverage, they are discussed in Appendix \ref{app:outputorich}.
In our notation, the prefix `G' (grain) denotes ice species.

The S- and Si-bearing refractory organic species originate from accretion of gas-phase species and their subsequent photoprocessing only.
This could be caused by the limited grain surface chemistry involving S and Si in the network.
The S-bearing ices are limited to GC$_n$S ($n = 2,3,4$), the Si-bearing ices to GSiC$_n$ ($n = 2,3,4$) and their hydrides.
The fractional contribution of S-bearing species increases with decreasing outflow density, as this shifts the onset of photodissociation closer to the star, increasing the formation of gas-phase daughter species of the parent CS.
The Si-bearing component increases with outflow density, but remains low ($\lesssim 0.1\%$).
This is mainly due to the low binding energy of the parent SiC$_2$ (1300 K).
The increase with outflow density is now caused by the shift of the onset photodissociation further away from the star, where SiC$_2$ is less likely to thermally desorb upon accretion.

All O-bearing refractory organic species are solely formed in grain surface reactions, initiated by the photodissociation of ice species on the grain surface.
Photodissociation of GH$_2$O results in GO and GOH, two radicals that readily react.
The main O-bearing component is formed via reactions with other simple radicals, such as GC$_n$ ($n = 1,2,3$). 
More complex species are formed via a pathway of reactions involving complex ices and radicals, leading to refractory organic material composed of two C atoms and up to two O atoms.
The fraction of O-bearing refractory organic material increases with outflow density because of the larger GH$_2$O abundance, caused by the larger outflow density as well as the delay in photodissociation of H$_2$O.
The increase in the O-bearing fraction leads to the increase in the O/C ratio of the refractory organic material.

The N-bearing refractory organic species are formed by accretion and subsequent photoprocessing as well as grain surface reactions.
Initially, GHC$_n$N ($n = 3,5,7,9$) are formed through accretion.
After the onset of photon-reactions, they can be photoprocessed or photodissociated into GC$_n$N, which can either reform GHC$_n$N or react with accreted GCH$_3$ to form GCH$_3$C$_n$N ($n = 3,5,7$).
These are the main N-bearing refractory organic species.
Others are formed via accretion, with possible hydration on the surface, followed by photoprocessing.
The fraction of N-bearing species increases with outflow density. 
This is linked to the gas-phase formation of HC$_n$N.
Because of their high binding energies (4580 K, 6180 K, 7780 K, 9380 K for $n = 3,5,7,9$), they are readily depleted onto the dust close to the star for all outflow densities.
Their peak gas-phase abundance shifts outwards with outflow density.
This leads to a much broader peak in GHC$_n$N for higher density outflows, as the gas-phase HC$_n$N are already photodissociated in the lower density outflows, and therefore to a richer N-bearing grain-surface chemistry.

The fraction of H-bearing refractory organic species increases with outflow density.
This is partly due to the increase with outflow density in N-bearing refractory organic species  and the decrease with outflow density in S-bearing refractory organic species, which do not bear any H.
The most important factor is the influence of the onset of photon-processes on the abundance of C-chains, with zero, one or two H atoms. 
In lower density outflows, the abundances of complex gas-phase daughter species increase closer to the star.
This leads to a larger abundance of their ices, which are readily photoprocessed into refractory organic material, removing the possibility of hydrogenation on the grain surface and leading to a smaller H-bearing component.

\subsection{Refractory organic contribution of AGB stars to the ISM and impact on other environments}			\label{subsect:discussion:output}

AGB stars are among the main contributors of dust and gas to the ISM, along with supernovae.
C-rich AGB stars are estimated to inject dust into the Milky Way at a rate of 3 M$_\odot$ pc$^{-2}$ Myr$^{-1}$, which corresponds to 53\% of the total C-rich dust injected.
Similarly, O-rich AGB stars have an estimated silicate dust injection rate of 5 M$_\odot$ pc$^{-2}$ Myr$^{-1}$, corresponding to 29\% of the total silicate dust injected \citep{Tielens2005}.
While we find that dust from O-rich outflows is covered by a negligible amount of refractory organic material, C-rich dust is on average covered by $3-9\%$ of such material, depending on the parent species used.
This implies that 53\% of the C-rich dust in the Milky Way does not enter the ISM completely bare, but has some refractory organic material on its surface.
However, the small average refractory organic coverage is unlikely to affect the physics and chemistry of the pristine dust entering the ISM.
Using the parent species of Papers I and II and the observed parents with maximal C-bearing species, 5\% and 14\% of the outflows, respectively, have a refractory organic coverage of at least 10\%.
The maximal coverage is 8\% when using the mean observed parent species.
Only in these high density outflows could the refractory organic coverage be able to significantly impact the dust's behaviour in the ISM surrounding the AGB star.
However, further research is needed to fully quantify its possible impact.

Based on observational searches for refractory organic material, refractory organic material in the diffuse ISM is mainly composed of aromatic and aliphatic hydrocarbons with a minimal oxygen content, corresponding to a O/C ratio $\sim$ 0.015 \citep{Pendleton2002}.
The O-bearing component of the refractory organic material on C-rich dust is very low (Fig. \ref{fig:outputobscrich-allfracs}). 
On average, over all outflows and all three sets of parent species, O only makes up $6.12 \times10^{-4}$ of the refractory organic material, corresponding to a O/C ratio of $\sim 9 \times 10^{-4}$, i.e., very depleted in oxygen.
Therefore, the refractory organic contribution of C-rich AGB stars is likely similar to that found in the diffuse ISM, making it an unlikely carrier of unidentified depleted oxygen.

Based on observations of the $3.4\ \mu$m spectral band in the diffuse ISM, at least $3-4\%$ of interstellar carbon is incorporated in aliphatic hydrocarbons in the refractory organic mantle.
These aliphatic hydrocarbons are attached to larger chemical structures, which likely contain up to 10\% of the interstellar carbon \citep{Pendleton2002}.
We find that only about $3 \times 10^{-7} - 2 \times 10^{-3}$ of the total C atoms in the outflow are incorporated into refractory organic material at the end of C-rich outflows.
This decreases to $5 \times 10^{-11} - 2 \times 10^{-7}$ for O-rich outflows.
Together with the low coverage of the AGB dust, it is clear that AGB outflows do not contribute significantly to the refractory organic mantle of interstellar dust.

\subsection{Sub-monolayer grain surface chemistry}			\label{subsect:discussion:surface}

Surface chemistry on dust within AGB outflows generally takes place in the sub-monolayer regime.
All outflows have less than one monolayer of ice on the grain surface, except higher density O-rich outflows which are covered by several layers of H$_2$O ice (Figs. \ref{fig:outputcrich} and \ref{fig:outputorich}).
Sub-monolayer regimes are not common in astrochemical environments, holding only in the transition region from diffuse to dense interstellar clouds, where ice mantles are formed, and during the desorption of ice mantles at snowlines in protoplanetary disks \citep{Potapov2019}.
AGB outflows can now be added to this list.

Compared to dust grains covered with ices, bare dust surfaces can lead to an increased reactivity for reactions involving molecules and/or radicals \citep{Potapov2020}.
This surface catalytic effect is likely due to the participation of the dust surface itself in the reactions and an increased diffusion rate
\citep{Potapov2017,Potapov2019,Potapov2020}.

Since the binding energy determines the competition between desorption and diffusion on the grain surface, we have simulated the possible influence of the surface catalytic effect by increasing the binding energies of all ice species in our network by 10\% and 30\%.
This simplified approach does not take into account the particular dust composition and surface properties or any species-specific increases.
Nonetheless, as a first approximation, we find that increasing the binding energy by 10\% and 30\% increases the refractory organic coverage by at least the same factor. 
The increase is larger for lower density outflows.
Compiling surface-dependent binding energies measured in the laboratory would be the next step in assessing the influence of surface catalytic effects.

\subsection{Sensitivity to model parameters}			\label{subsect:discussion:uncert}

Several assumptions have been made in order to calculate the observational grids of O-rich and C-rich outflows.
Here, we address their possible impact on the formation of refractory organic material.

\begin{figure}
 \includegraphics[width=1.\columnwidth]{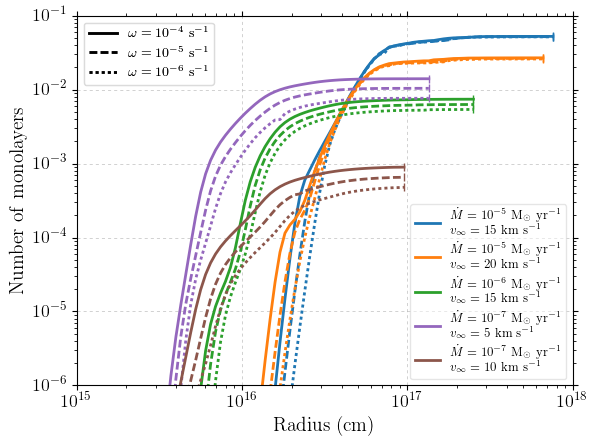}
 \caption{Number of monolayers of refractory organic material in C-rich outflows with different mass-loss rates and outflow velocities, using the parent species of Papers I and II.
 The different line styles correspond to different values of the photoprocessing rate, $\omega$.
 The dash marks where the H$_2$ number density reaches 10 cm$^{-3}$, i.e., where the outflow merges with the ISM.
  }
 \label{fig:effectrate}
\end{figure}

\paragraph*{Photoprocessing rate}			\label{subsect:discussion:uncert:rate}

We assumed a photoprocessing rate of $\omega =10^{-5}$ s$^{-1}$, which lies within the range of $3 \times 10^{-4} - 8 \times 10^{-6}$ s$^{-1}$ derived from irradiation times of laboratory experiments \citep{Allamandola1988,Bernstein2002}.
The effect of changing $\omega$ within this range is shown in Fig. \ref{fig:effectrate} for different C-rich outflows, using the parent species of Papers I and II (as in  Fig. \ref{fig:outputcrich}).
Only lower density outflows show a significant effect when changing $\omega$.
When varying the rate between $10^{-4}$ and $10^{-6}$ s$^{-1}$, the refractory organic output in the lowest density outflows varies by a factor of about two. 
This range decreases with increasing outflow density and becomes negligible for the highest density outflows. 
As higher density outflows are the main carriers of refractory organic material, changing the photoprocessing rate within the range derived from experiments will not alter our main conclusions.

\paragraph*{Grain size distribution}			\label{subsect:discussion:uncert:gsd}

The assumed GSD of the dust influences the formation of the ice mantle (Paper II).
While assuming that the GSD has been set in the inner wind and the grains do not grow throughout the subsequent outflow, a GSD with a larger average grain surface area leads to a larger depletion of gas-phase species and therefore to a thicker ice mantle.
Observations of dust grains around AGB stars suggest that AGB dust is large but not single-sized \cite[e.g.,][]{Groenewegen1997,Winters1997,Hoppe2000,Khouri2016,DellAgli2017,Nanni2018}.
In Paper II, we found that molecular line observations tentatively suggest an average grain surface area larger than that of the canonical MRN distribution.

A larger average grain surface would lead to a thicker ice mantle and subsequently more refractory organic material.
However, because the precise GSD is not yet known, we have assumed the commonly-assumed canonical MRN distribution.
In order to quantify the GSD of AGB dust, further observations are necessary as well as a more complete understanding of the dust formation process(es).
One key aspect for the formation of refractory organic material is the possible dependence of the GSD on the outflow density.

\paragraph*{Drift velocity}			\label{subsect:discussion:uncert:drift}

The drift velocity between dust and gas has a large impact on the accretion of gas-phase species.
Higher drift velocities lead to more accretion, and hence to a larger ice mantle, but only up to $v_\mathrm{drift} \sim 10$ km s$^{-1}$.  
For larger values, sputtering partly destroys the ice mantle (Paper I).

When modelling the observational grids, we assumed a drift velocity based on the observed mass-loss rate, as higher mass-loss rate outflows generally have lower drift velocities.
Using drift velocities specific to the outflow, retrieved through a combination of continuum and line emission radiative transfer, will make the estimate of the refractory organic coverage more accurate.

\paragraph*{Spherical symmetry}			\label{subsect:discussion:uncert:dust}

A central assumption to the chemical model is the spherical symmetry of the outflow.
However, both large-scale and small-scale deviations from spherical symmetry have been observed, such as spirals \cite[e.g.,][]{Mauron2006,Maercker2012}, equatorial density enhancements (EDEs) or disks \citep{Kervella2014,Homan2018} and clumps \cite[e.g.][]{Khouri2016,Agundez2017}.
The presence of overdensities could lead to an increase in ice formation in these regions.
Models specific to such density distributions are necessary to determine whether this leads to a larger refractory organic coverage.

In particular, the presence of an EDE can lead to the overestimation of the mass-loss rate \citep{Decin2019}.
Hydrodynamical models suggest that the overestimation can be a factor of a few \citep{ElMellah2020}. 
A smaller mass-loss rate would imply a smaller refractory organic coverage.
However, the density within EDE can be up to several orders of magnitude larger than that of the surrounding outflow, depending on the intrinsic orbital parameters of the binary system and its inclination.
Such an increase in density would have a positive effect on refractory organic formation.
To disentangle the effects of the decrease in mass-loss rate and the overdensity within the EDE, specialised (physico)chemical models are needed.

Moreover, large-scale density structures such as EDEs and spirals are likely due to binary interaction of the AGB star with a stellar or planetary companion \cite[e.g.,][]{Mauron2006,Kervella2014,Ramstedt2014,Kim2015,Decin2015,Maercker2016,Decin2020}.
If the binary companion has a strong UV field, e.g., a white dwarf, photon-processes are able to occur in the otherwise shielded inner region.
Besides the impact on the gas-phase chemistry \citep{VandeSande2019a}, this could initiate grain-surface chemistry as well as photoprocessing.
The further development of chemical models that account for inner UV photons is necessary.

\section{Conclusions}				\label{sect:conclusions}

We presented the first observationally motivated study of the refractory organic contribution of AGB outflows to the ISM.
The chemical network of Papers I and II was expanded to include a more complex grain surface chemistry, reactive desorption and photodissociation on the grain surface.
A new type of reaction was introduced based on laboratory experiments: the photoprocessing of volatile complex ices into inert refractory organic material.
This reaction is a sink for complex ices, i.e., ices containing at least three C atoms or two C atoms and one other heavy atom.
We assembled grids of O- and C-rich outflows from observational samples.
New observationally motivated C-rich parent species were also compiled, with one representing an average outflow and the other an outflow with the maximal observed carbon content.

Refractory organic material is able to form on the dust within the outflow.
The majority of the material is inherited directly from the gas phase: complex gas-phase species are accreted onto the dust and subsequently photoprocessed into refractory organic material.
Grain-surface chemistry plays only a minor role in the formation of complex ice species.
This is unlike other astrochemical environments, where reactions between simple ices are crucial to chemical complexity.
The amount of refractory organic material formed depends on the outflow density and the initial gas-phase composition.
The outflow density plays a twofold role, as it sets the accretion rate and the onset of photon-processes (photodissociation in the gas and on the ice, photodesorption and photoprocessing).
Larger density outflows have a larger ice mantle and refractory organic coverage.
The initial gas-phase composition determines the availability of feedstock ices.
The initial abundances of C$_2$H$_2$ and, to a lesser extent, HCN are crucial to the refractory organic coverage.

Therefore, dust in O-rich outflows is only negligibly covered by refractory organic material.
The average refractory organic coverage is only 0.0001\%.
In C-rich outflow, the dust has an average refractory organic coverage of $3-9\%$, depending on the parent species used.
This corresponds to only about $10^{-7} - 10^{-3}$ of all C atoms being incorporated into the refractory organic mantle.
The coverage increases with increasing outflow density up to $8-22\%$.
However, the small average refractory organic coverage is unlikely to affect the physics and chemistry of the dust in the ISM.
The oxygen content of the material agrees with the low O/C ratios of $\sim 0.015$ measured in the ISM \citep{Pendleton2002}.

In order to improve our estimate of the refractory organic contribution to the ISM, measured drift velocities and a better grasp on the GSD of the dust are necessary.
These two parameters influence the accretion rate of gas-phase species.
Changing the photoprocessing rate within the range obtained from laboratory experiments does not significantly affect our results.
The grain-surface chemistry takes place in a sub-monolayer regime, which is not common.
The surface catalytic effect could increase the refractory organic coverage, but further laboratory experiments using bare dust grains are necessary.
Additionally, our spherically symmetric model cannot account for the possible effects of small- and large-scale overdensities within the outflow.
For these outflows, models specific to the particular density distribution have to be designed.
The impact of a UV-emitting binary companion could also influence the refractory organic output.

This work shows that photoprocessing is a viable chemical process in AGB outflows.
Although refractory organic material can be formed in specific outflows, it does not significantly contribute to the refractory organic mantles of interstellar dust.
It opens up the question to the refractory organic coverage of other contributors to dust in the ISM, such as supernovae.
Characterising pristine stellar dust is not only crucial to understanding its subsequent evolution in the ISM, but also to understanding the origin of refractory organics on interstellar dust.

\section*{Acknowledgements}

MVdS acknowledges support from the Research Foundation Flanders (FWO) through grant 12X6419N.
CW acknowledges financial support from the University of Leeds and from the Science and Technology Facilities Council (grant numbers ST/R000549/1 and ST/T000287/1).
TJM thanks STFC for support under grant reference ST/P000312/1.

\section*{Data availability}

The chemical reaction network, chemical kinetics model and all data underlying this article are available on request to the corresponding author.




\bibliographystyle{mnras}
\bibliography{chemistry} 

\begin{thebibliography}{}
\makeatletter
\relax
\def\mn@urlcharsother{\let\do\@makeother \do\$\do\&\do\#\do\^\do\_\do\%\do\~}
\def\mn@doi{\begingroup\mn@urlcharsother \@ifnextchar [ {\mn@doi@}
  {\mn@doi@[]}}
\def\mn@doi@[#1]#2{\def\@tempa{#1}\ifx\@tempa\@empty \href
  {http://dx.doi.org/#2} {doi:#2}\else \href {http://dx.doi.org/#2} {#1}\fi
  \endgroup}
\def\mn@eprint#1#2{\mn@eprint@#1:#2::\@nil}
\def\mn@eprint@arXiv#1{\href {http://arxiv.org/abs/#1} {{\tt arXiv:#1}}}
\def\mn@eprint@dblp#1{\href {http://dblp.uni-trier.de/rec/bibtex/#1.xml}
  {dblp:#1}}
\def\mn@eprint@#1:#2:#3:#4\@nil{\def\@tempa {#1}\def\@tempb {#2}\def\@tempc
  {#3}\ifx \@tempc \@empty \let \@tempc \@tempb \let \@tempb \@tempa \fi \ifx
  \@tempb \@empty \def\@tempb {arXiv}\fi \@ifundefined
  {mn@eprint@\@tempb}{\@tempb:\@tempc}{\expandafter \expandafter \csname
  mn@eprint@\@tempb\endcsname \expandafter{\@tempc}}}

\bibitem[\protect\citeauthoryear{{Ag{\'u}ndez} et~al.,}{{Ag{\'u}ndez}
  et~al.}{2017}]{Agundez2017}
{Ag{\'u}ndez} M.,  et~al., 2017, \mn@doi [\aap] {10.1051/0004-6361/201630274},
  \href {http://adsabs.harvard.edu/abs/2017A%26A...601A...4A} {601, A4}

\bibitem[\protect\citeauthoryear{{Ag{\'u}ndez}, {Mart{\'\i}nez}, {de Andres},
  {Cernicharo}  \& {Mart{\'\i}n-Gago}}{{Ag{\'u}ndez}
  et~al.}{2020}]{Agundez2020}
{Ag{\'u}ndez} M.,  {Mart{\'\i}nez} J.~I.,  {de Andres} P.~L.,  {Cernicharo} J.,
    {Mart{\'\i}n-Gago} J.~A.,  2020, \mn@doi [\aap]
  {10.1051/0004-6361/202037496}, \href
  {https://ui.adsabs.harvard.edu/abs/2020A&A...637A..59A} {637, A59}

\bibitem[\protect\citeauthoryear{Allamandola, Sandford  \& Valero}{Allamandola
  et~al.}{1988}]{Allamandola1988}
Allamandola L.,  Sandford S.,   Valero G.,  1988, \mn@doi [Icarus]
  {https://doi.org/10.1016/0019-1035(88)90070-X}, 76, 225

\bibitem[\protect\citeauthoryear{{Bernstein}, {Dworkin}, {Sandford}, {Cooper}
  \& {Allamandola}}{{Bernstein} et~al.}{2002}]{Bernstein2002}
{Bernstein} M.~P.,  {Dworkin} J.~P.,  {Sandford} S.~A.,  {Cooper} G.~W.,
  {Allamandola} L.~J.,  2002, \mn@doi [\nat] {10.1038/416401a}, \href
  {https://ui.adsabs.harvard.edu/abs/2002Natur.416..401B} {416, 401}

\bibitem[\protect\citeauthoryear{{Bischoff}, {Kreuzig}, {Haack}, {Gundlach}  \&
  {Blum}}{{Bischoff} et~al.}{2020}]{Bischoff2020}
{Bischoff} D.,  {Kreuzig} C.,  {Haack} D.,  {Gundlach} B.,   {Blum} J.,  2020,
  \mn@doi [\mnras] {10.1093/mnras/staa2126}, \href
  {https://ui.adsabs.harvard.edu/abs/2020MNRAS.497.2517B} {497, 2517}

\bibitem[\protect\citeauthoryear{{Butchart}, {McFadzean}, {Whittet}, {Geballe}
  \& {Greenberg}}{{Butchart} et~al.}{1986}]{Butchart1986}
{Butchart} I.,  {McFadzean} A.~D.,  {Whittet} D.~C.~B.,  {Geballe} T.~R.,
  {Greenberg} J.~M.,  1986, \aap, \href
  {https://ui.adsabs.harvard.edu/abs/1986A&A...154L...5B} {154, L5}

\bibitem[\protect\citeauthoryear{{Cuppen}, {Walsh}, {Lamberts}, {Semenov},
  {Garrod}, {Penteado}  \& {Ioppolo}}{{Cuppen} et~al.}{2017}]{Cuppen2017}
{Cuppen} H.~M.,  {Walsh} C.,  {Lamberts} T.,  {Semenov} D.,  {Garrod} R.~T.,
  {Penteado} E.~M.,   {Ioppolo} S.,  2017, \mn@doi [\ssr]
  {10.1007/s11214-016-0319-3}, \href
  {http://adsabs.harvard.edu/abs/2017SSRv..212....1C} {212, 1}

\bibitem[\protect\citeauthoryear{{Danilovich} et~al.,}{{Danilovich}
  et~al.}{2015}]{Danilovich2015}
{Danilovich} T.,  et~al., 2015, \mn@doi [\aap] {10.1051/0004-6361/201526705},
  \href {https://ui.adsabs.harvard.edu/abs/2015A&A...581A..60D} {581, A60}

\bibitem[\protect\citeauthoryear{{Decin} et~al.,}{{Decin}
  et~al.}{2010}]{Decin2010}
{Decin} L.,  et~al., 2010, \mn@doi [\aap] {10.1051/0004-6361/201014136}, \href
  {http://adsabs.harvard.edu/abs/2010A%26A...516A..69D} {516, A69}

\bibitem[\protect\citeauthoryear{{Decin}, {Richards}, {Neufeld}, {Steffen},
  {Melnick}  \& {Lombaert}}{{Decin} et~al.}{2015}]{Decin2015}
{Decin} L.,  {Richards} A.~M.~S.,  {Neufeld} D.,  {Steffen} W.,  {Melnick} G.,
   {Lombaert} R.,  2015, \mn@doi [\aap] {10.1051/0004-6361/201424593}, \href
  {https://ui.adsabs.harvard.edu/abs/2015A&A...574A...5D} {574, A5}

\bibitem[\protect\citeauthoryear{{Decin} et~al.,}{{Decin}
  et~al.}{2019}]{Decin2019}
{Decin} L.,  et~al., 2019, \mn@doi [Nature Astronomy]
  {10.1038/s41550-019-0703-5}, \href
  {https://ui.adsabs.harvard.edu/abs/2019NatAs...3..408D} {3, 408}

\bibitem[\protect\citeauthoryear{Decin et~al.,}{Decin et~al.}{2020}]{Decin2020}
Decin L.,  et~al., 2020, \mn@doi [Science] {10.1126/science.abb1229}, 369, 1497

\bibitem[\protect\citeauthoryear{{Dell'Agli}, {Garc{\'\i}a-Hern{\'a}ndez},
  {Schneider}, {Ventura}, {La Franca}, {Valiante}, {Marini}  \& {Di
  Criscienzo}}{{Dell'Agli} et~al.}{2017}]{DellAgli2017}
{Dell'Agli} F.,  {Garc{\'\i}a-Hern{\'a}ndez} D.~A.,  {Schneider} R.,  {Ventura}
  P.,  {La Franca} F.,  {Valiante} R.,  {Marini} E.,   {Di Criscienzo} M.,
  2017, \mn@doi [\mnras] {10.1093/mnras/stx387}, \href
  {https://ui.adsabs.harvard.edu/abs/2017MNRAS.467.4431D} {467, 4431}

\bibitem[\protect\citeauthoryear{{Draine}}{{Draine}}{2003}]{Draine2003}
{Draine} B.~T.,  2003, \mn@doi [\araa]
  {10.1146/annurev.astro.41.011802.094840}, \href
  {http://adsabs.harvard.edu/abs/2003ARA%26A..41..241D} {41, 241}

\bibitem[\protect\citeauthoryear{{Duley}, {Scott}, {Seahra}  \&
  {Dadswell}}{{Duley} et~al.}{1998}]{Duley1998}
{Duley} W.~W.,  {Scott} A.~D.,  {Seahra} S.,   {Dadswell} G.,  1998, \mn@doi
  [\apjl] {10.1086/311548}, \href
  {https://ui.adsabs.harvard.edu/abs/1998ApJ...503L.183D} {503, L183}

\bibitem[\protect\citeauthoryear{{El Mellah}, {Bolte}, {Decin}, {Homan}  \&
  {Keppens}}{{El Mellah} et~al.}{2020}]{ElMellah2020}
{El Mellah} I.,  {Bolte} J.,  {Decin} L.,  {Homan} W.,   {Keppens} R.,  2020,
  \mn@doi [\aap] {10.1051/0004-6361/202037492}, \href
  {https://ui.adsabs.harvard.edu/abs/2020A&A...637A..91E} {637, A91}

\bibitem[\protect\citeauthoryear{{Garrod}, {Widicus Weaver}  \&
  {Herbst}}{{Garrod} et~al.}{2008}]{Garrod2008}
{Garrod} R.~T.,  {Widicus Weaver} S.~L.,   {Herbst} E.,  2008, \mn@doi [\apj]
  {10.1086/588035}, \href
  {https://ui.adsabs.harvard.edu/abs/2008ApJ...682..283G} {682, 283}

\bibitem[\protect\citeauthoryear{{Gonz{\'a}lez Delgado}, {Olofsson},
  {Kerschbaum}, {Sch{\"o}ier}, {Lindqvist}  \& {Groenewegen}}{{Gonz{\'a}lez
  Delgado} et~al.}{2003}]{GonzalezDelgado2003}
{Gonz{\'a}lez Delgado} D.,  {Olofsson} H.,  {Kerschbaum} F.,  {Sch{\"o}ier}
  F.~L.,  {Lindqvist} M.,   {Groenewegen} M.~A.~T.,  2003, \mn@doi [\aap]
  {10.1051/0004-6361:20031068}, \href
  {http://adsabs.harvard.edu/abs/2003A%26A...411..123G} {411, 123}

\bibitem[\protect\citeauthoryear{{Greenberg}}{{Greenberg}}{1986}]{Greenberg1986}
{Greenberg} J.~M.,  1986, \mn@doi [\apss] {10.1007/BF00656015}, \href
  {https://ui.adsabs.harvard.edu/abs/1986Ap&SS.128...17G} {128, 17}

\bibitem[\protect\citeauthoryear{{Greenberg} \& {Shen}}{{Greenberg} \&
  {Shen}}{1999}]{Greenberg1999}
{Greenberg} J.~M.,  {Shen} C.,  1999, \mn@doi [\apss]
  {10.1023/A:1017003503904}, \href
  {https://ui.adsabs.harvard.edu/abs/1999Ap&SS.269...33G} {269, 33}

\bibitem[\protect\citeauthoryear{{Greenberg}, {Yencha}, {Corbett}  \&
  {Frisch}}{{Greenberg} et~al.}{1972}]{Greenberg1972}
{Greenberg} J.~M.,  {Yencha} A.~J.,  {Corbett} J.~W.,   {Frisch} H.~L.,  1972,
  in Les Spectres des Astres dans l'Infrarouge et les Microondes. pp 425--436

\bibitem[\protect\citeauthoryear{{Groenewegen}}{{Groenewegen}}{1997}]{Groenewegen1997}
{Groenewegen} M.~A.~T.,  1997, \aap, \href
  {https://ui.adsabs.harvard.edu/abs/1997A&A...317..503G} {317, 503}

\bibitem[\protect\citeauthoryear{{Hagen}, {Allamandola}  \&
  {Greenberg}}{{Hagen} et~al.}{1979}]{Hagen1979}
{Hagen} W.,  {Allamandola} L.~J.,   {Greenberg} J.~M.,  1979, \mn@doi [\apss]
  {10.1007/BF00643502}, \href
  {https://ui.adsabs.harvard.edu/abs/1979Ap&SS..65..215H} {65, 215}

\bibitem[\protect\citeauthoryear{{Henning} \& {Semenov}}{{Henning} \&
  {Semenov}}{2013}]{Henning2013}
{Henning} T.,  {Semenov} D.,  2013, \mn@doi [Chemical Reviews]
  {10.1021/cr400128p}, \href
  {https://ui.adsabs.harvard.edu/abs/2013ChRv..113.9016H} {113, 9016}

\bibitem[\protect\citeauthoryear{{Heras} \& {Hony}}{{Heras} \&
  {Hony}}{2005}]{Heras2005}
{Heras} A.~M.,  {Hony} S.,  2005, \mn@doi [\aap] {10.1051/0004-6361:20042296},
  \href {http://adsabs.harvard.edu/abs/2005A%26A...439..171H} {439, 171}

\bibitem[\protect\citeauthoryear{{H{\"o}fner} \& {Olofsson}}{{H{\"o}fner} \&
  {Olofsson}}{2018}]{Hofner2018}
{H{\"o}fner} S.,  {Olofsson} H.,  2018, \mn@doi [\aapr]
  {10.1007/s00159-017-0106-5}, \href
  {https://ui.adsabs.harvard.edu/abs/2018A&ARv..26....1H} {26, 1}

\bibitem[\protect\citeauthoryear{{Homan}, {Danilovich}, {Decin}, {de Koter},
  {Nuth}  \& {Van de Sande}}{{Homan} et~al.}{2018}]{Homan2018}
{Homan} W.,  {Danilovich} T.,  {Decin} L.,  {de Koter} A.,  {Nuth} J.,   {Van
  de Sande} M.,  2018, \mn@doi [\aap] {10.1051/0004-6361/201732246}, \href
  {https://ui.adsabs.harvard.edu/abs/2018A&A...614A.113H} {614, A113}

\bibitem[\protect\citeauthoryear{Hoppe \& Zinner}{Hoppe \&
  Zinner}{2000}]{Hoppe2000}
Hoppe P.,  Zinner E.,  2000, \mn@doi [Journal of Geophysical Research: Space
  Physics] {10.1029/1999JA900194}, 105, 10371

\bibitem[\protect\citeauthoryear{{J{\"a}ger}, {Dorschner}, {Mutschke}, {Posch}
  \& {Henning}}{{J{\"a}ger} et~al.}{2003}]{Jager2003}
{J{\"a}ger} C.,  {Dorschner} J.,  {Mutschke} H.,  {Posch} T.,   {Henning} T.,
  2003, \mn@doi [\aap] {10.1051/0004-6361:20030916}, \href
  {https://ui.adsabs.harvard.edu/abs/2003A%26A...408..193J} {408, 193}

\bibitem[\protect\citeauthoryear{{Jones}, {Fanciullo}, {K{\"o}hler},
  {Verstraete}, {Guillet}, {Bocchio}  \& {Ysard}}{{Jones}
  et~al.}{2013}]{Jones2013}
{Jones} A.~P.,  {Fanciullo} L.,  {K{\"o}hler} M.,  {Verstraete} L.,  {Guillet}
  V.,  {Bocchio} M.,   {Ysard} N.,  2013, \mn@doi [\aap]
  {10.1051/0004-6361/201321686}, \href
  {https://ui.adsabs.harvard.edu/abs/2013A&A...558A..62J} {558, A62}

\bibitem[\protect\citeauthoryear{{Kervella} et~al.,}{{Kervella}
  et~al.}{2014}]{Kervella2014}
{Kervella} P.,  et~al., 2014, \mn@doi [\aap] {10.1051/0004-6361/201323273},
  \href {http://adsabs.harvard.edu/abs/2014A%26A...564A..88K} {564, A88}

\bibitem[\protect\citeauthoryear{{Khouri} et~al.,}{{Khouri}
  et~al.}{2016}]{Khouri2016}
{Khouri} T.,  et~al., 2016, \mn@doi [\aap] {10.1051/0004-6361/201628435}, \href
  {http://adsabs.harvard.edu/abs/2016A%26A...591A..70K} {591, A70}

\bibitem[\protect\citeauthoryear{{Kim} et~al.,}{{Kim} et~al.}{2015}]{Kim2015}
{Kim} H.,  et~al., 2015, \mn@doi [\apj] {10.1088/0004-637X/814/1/61}, \href
  {https://ui.adsabs.harvard.edu/abs/2015ApJ...814...61K} {814, 61}

\bibitem[\protect\citeauthoryear{{Kouchi}, {Kudo}, {Nakano}, {Arakawa},
  {Watanabe}, {Sirono}, {Higa}  \& {Maeno}}{{Kouchi} et~al.}{2002}]{Kouchi2002}
{Kouchi} A.,  {Kudo} T.,  {Nakano} H.,  {Arakawa} M.,  {Watanabe} N.,  {Sirono}
  S.-i.,  {Higa} M.,   {Maeno} N.,  2002, \mn@doi [\apjl] {10.1086/339618},
  \href {https://ui.adsabs.harvard.edu/abs/2002ApJ...566L.121K} {566, L121}

\bibitem[\protect\citeauthoryear{{Maercker} et~al.,}{{Maercker}
  et~al.}{2012}]{Maercker2012}
{Maercker} M.,  et~al., 2012, \mn@doi [\nat] {10.1038/nature11511}, \href
  {http://adsabs.harvard.edu/abs/2012Natur.490..232M} {490, 232}

\bibitem[\protect\citeauthoryear{{Maercker}, {Danilovich}, {Olofsson}, {De
  Beck}, {Justtanont}, {Lombaert}  \& {Royer}}{{Maercker}
  et~al.}{2016}]{Maercker2016}
{Maercker} M.,  {Danilovich} T.,  {Olofsson} H.,  {De Beck} E.,  {Justtanont}
  K.,  {Lombaert} R.,   {Royer} P.,  2016, \mn@doi [\aap]
  {10.1051/0004-6361/201628310}, \href
  {http://adsabs.harvard.edu/abs/2016A%26A...591A..44M} {591, A44}

\bibitem[\protect\citeauthoryear{{Materese}, {Nuevo}  \& {Sandford}}{{Materese}
  et~al.}{2017}]{Materese2017}
{Materese} C.~K.,  {Nuevo} M.,   {Sandford} S.~A.,  2017, \mn@doi
  [Astrobiology] {10.1089/ast.2016.1613}, \href
  {https://ui.adsabs.harvard.edu/abs/2017AsBio..17..761M} {17, 761}

\bibitem[\protect\citeauthoryear{{Mathis}}{{Mathis}}{1979}]{Mathis1979}
{Mathis} J.~S.,  1979, \mn@doi [\apj] {10.1086/157335}, \href
  {https://ui.adsabs.harvard.edu/abs/1979ApJ...232..747M} {232, 747}

\bibitem[\protect\citeauthoryear{{Mathis}, {Rumpl}  \& {Nordsieck}}{{Mathis}
  et~al.}{1977}]{Mathis1977}
{Mathis} J.~S.,  {Rumpl} W.,   {Nordsieck} K.~H.,  1977, \mn@doi [\apj]
  {10.1086/155591}, \href {http://adsabs.harvard.edu/abs/1977ApJ...217..425M}
  {217, 425}

\bibitem[\protect\citeauthoryear{{Mauron} \& {Huggins}}{{Mauron} \&
  {Huggins}}{2006}]{Mauron2006}
{Mauron} N.,  {Huggins} P.~J.,  2006, \mn@doi [\aap]
  {10.1051/0004-6361:20054739}, \href
  {http://adsabs.harvard.edu/abs/2006A%26A...452..257M} {452, 257}

\bibitem[\protect\citeauthoryear{{McElroy}, {Walsh}, {Markwick}, {Cordiner},
  {Smith}  \& {Millar}}{{McElroy} et~al.}{2013}]{McElroy2013}
{McElroy} D.,  {Walsh} C.,  {Markwick} A.~J.,  {Cordiner} M.~A.,  {Smith} K.,
  {Millar} T.~J.,  2013, \mn@doi [\aap] {10.1051/0004-6361/201220465}, \href
  {http://adsabs.harvard.edu/abs/2013A%26A...550A..36M} {550, A36}

\bibitem[\protect\citeauthoryear{{Millar} \& {Herbst}}{{Millar} \&
  {Herbst}}{1994}]{Millar1994}
{Millar} T.~J.,  {Herbst} E.,  1994, \aap, \href
  {http://adsabs.harvard.edu/abs/1994A%26A...288..561M} {288, 561}

\bibitem[\protect\citeauthoryear{{Min}, {Dullemond}, {Dominik}, {de Koter}  \&
  {Hovenier}}{{Min} et~al.}{2009}]{Min2009}
{Min} M.,  {Dullemond} C.~P.,  {Dominik} C.,  {de Koter} A.,   {Hovenier}
  J.~W.,  2009, \mn@doi [\aap] {10.1051/0004-6361/200811470}, \href
  {http://adsabs.harvard.edu/abs/2009A%26A...497..155M} {497, 155}

\bibitem[\protect\citeauthoryear{{Mu{\~n}oz Caro} \& {Schutte}}{{Mu{\~n}oz
  Caro} \& {Schutte}}{2003}]{MunozCaro2003}
{Mu{\~n}oz Caro} G.~M.,  {Schutte} W.~A.,  2003, \mn@doi [\aap]
  {10.1051/0004-6361:20031408}, \href
  {https://ui.adsabs.harvard.edu/abs/2003A&A...412..121M} {412, 121}

\bibitem[\protect\citeauthoryear{{Mutschke}, {Begemann}, {Dorschner},
  {Guertler}, {Gustafson}, {Henning}  \& {Stognienko}}{{Mutschke}
  et~al.}{1998}]{Mutschke1998}
{Mutschke} H.,  {Begemann} B.,  {Dorschner} J.,  {Guertler} J.,  {Gustafson}
  B.,  {Henning} T.,   {Stognienko} R.,  1998, \aap, \href
  {http://adsabs.harvard.edu/abs/1998A%26A...333..188M} {333, 188}

\bibitem[\protect\citeauthoryear{{Nanni}, {Marigo}, {Girardi}, {Rubele},
  {Bressan}, {Groenewegen}, {Pastorelli}  \& {Aringer}}{{Nanni}
  et~al.}{2018}]{Nanni2018}
{Nanni} A.,  {Marigo} P.,  {Girardi} L.,  {Rubele} S.,  {Bressan} A.,
  {Groenewegen} M. A.~T.,  {Pastorelli} G.,   {Aringer} B.,  2018, \mn@doi
  [\mnras] {10.1093/mnras/stx2641}, \href
  {https://ui.adsabs.harvard.edu/abs/2018MNRAS.473.5492N} {473, 5492}

\bibitem[\protect\citeauthoryear{{Olofsson}, {Gonz{\'a}lez Delgado},
  {Kerschbaum}  \& {Sch{\"o}ier}}{{Olofsson} et~al.}{2002}]{Olofsson2002}
{Olofsson} H.,  {Gonz{\'a}lez Delgado} D.,  {Kerschbaum} F.,   {Sch{\"o}ier}
  F.~L.,  2002, \mn@doi [\aap] {10.1051/0004-6361:20020841}, \href
  {http://adsabs.harvard.edu/abs/2002A%26A...391.1053O} {391, 1053}

\bibitem[\protect\citeauthoryear{{Pendleton} \& {Allamandola}}{{Pendleton} \&
  {Allamandola}}{2002}]{Pendleton2002}
{Pendleton} Y.~J.,  {Allamandola} L.~J.,  2002, \mn@doi [\apjs]
  {10.1086/322999}, \href
  {https://ui.adsabs.harvard.edu/abs/2002ApJS..138...75P} {138, 75}

\bibitem[\protect\citeauthoryear{Potapov, J\"ager, Henning, Jonusas  \&
  Krim}{Potapov et~al.}{2017}]{Potapov2017}
Potapov A.,  J\"ager C.,  Henning T.,  Jonusas M.,   Krim L.,  2017, \mn@doi
  [The Astrophysical Journal] {10.3847/1538-4357/aa85e8}, 846, 131

\bibitem[\protect\citeauthoryear{Potapov, Theul{\'{e}}, J\"ager  \&
  Henning}{Potapov et~al.}{2019}]{Potapov2019}
Potapov A.,  Theul{\'{e}} P.,  J\"ager C.,   Henning T.,  2019, \mn@doi [The
  Astrophysical Journal] {10.3847/2041-8213/ab2538}, 878, L20

\bibitem[\protect\citeauthoryear{{Potapov}, {J{\"a}ger}  \&
  {Henning}}{{Potapov} et~al.}{2020}]{Potapov2020}
{Potapov} A.,  {J{\"a}ger} C.,   {Henning} T.,  2020, arXiv e-prints, \href
  {https://ui.adsabs.harvard.edu/abs/2020arXiv200500757P} {p. arXiv:2005.00757}

\bibitem[\protect\citeauthoryear{{Preibisch}, {Ossenkopf}, {Yorke}  \&
  {Henning}}{{Preibisch} et~al.}{1993}]{Preibisch1993}
{Preibisch} T.,  {Ossenkopf} V.,  {Yorke} H.~W.,   {Henning} T.,  1993, \aap,
  \href {https://ui.adsabs.harvard.edu/abs/1993A%26A...279..577P} {279, 577}

\bibitem[\protect\citeauthoryear{{Ramstedt}, {Sch{\"o}ier}, {Olofsson}  \&
  {Lundgren}}{{Ramstedt} et~al.}{2008}]{Ramstedt2008}
{Ramstedt} S.,  {Sch{\"o}ier} F.~L.,  {Olofsson} H.,   {Lundgren} A.~A.,  2008,
  \mn@doi [\aap] {10.1051/0004-6361:20078876}, \href
  {https://ui.adsabs.harvard.edu/abs/2008A&A...487..645R} {487, 645}

\bibitem[\protect\citeauthoryear{{Ramstedt} et~al.,}{{Ramstedt}
  et~al.}{2014}]{Ramstedt2014}
{Ramstedt} S.,  et~al., 2014, \mn@doi [\aap] {10.1051/0004-6361/201425029},
  \href {https://ui.adsabs.harvard.edu/abs/2014A&A...570L..14R} {570, L14}

\bibitem[\protect\citeauthoryear{{Rau}, {Hron}, {Paladini}, {Aringer},
  {Eriksson}, {Marigo}, {Nowotny}  \& {Grellmann}}{{Rau}
  et~al.}{2017}]{Rau2017}
{Rau} G.,  {Hron} J.,  {Paladini} C.,  {Aringer} B.,  {Eriksson} K.,  {Marigo}
  P.,  {Nowotny} W.,   {Grellmann} R.,  2017, \mn@doi [\aap]
  {10.1051/0004-6361/201629337}, \href
  {https://ui.adsabs.harvard.edu/abs/2017A&A...600A..92R} {600, A92}

\bibitem[\protect\citeauthoryear{{Sandford}, {Allamandola}, {Tielens},
  {Sellgren}, {Tapia}  \& {Pendleton}}{{Sandford} et~al.}{1991}]{Sandford1991}
{Sandford} S.~A.,  {Allamandola} L.~J.,  {Tielens} A.~G.~G.~M.,  {Sellgren} K.,
   {Tapia} M.,   {Pendleton} Y.,  1991, \mn@doi [\apj] {10.1086/169925}, \href
  {https://ui.adsabs.harvard.edu/abs/1991ApJ...371..607S} {371, 607}

\bibitem[\protect\citeauthoryear{{Sch{\"o}ier} \& {Olofsson}}{{Sch{\"o}ier} \&
  {Olofsson}}{2001}]{Schoier2001}
{Sch{\"o}ier} F.~L.,  {Olofsson} H.,  2001, \mn@doi [\aap]
  {10.1051/0004-6361:20010072}, \href
  {https://ui.adsabs.harvard.edu/abs/2001A&A...368..969S} {368, 969}

\bibitem[\protect\citeauthoryear{{Tielens}}{{Tielens}}{2005}]{Tielens2005}
{Tielens} A.~G.~G.~M.,  2005, {The Physics and Chemistry of the Interstellar
  Medium}.
Cambridge University Press

\bibitem[\protect\citeauthoryear{Tielens}{Tielens}{2013}]{Tielens2013}
Tielens A. G. G.~M.,  2013, \mn@doi [Rev. Mod. Phys.]
  {10.1103/RevModPhys.85.1021}, 85, 1021

\bibitem[\protect\citeauthoryear{{Van de Sande} \& {Millar}}{{Van de Sande} \&
  {Millar}}{2019}]{VandeSande2019a}
{Van de Sande} M.,  {Millar} T.~J.,  2019, \mn@doi [\apj]
  {10.3847/1538-4357/ab03d4}, \href
  {https://ui.adsabs.harvard.edu/abs/2019ApJ...873...36V} {873, 36}

\bibitem[\protect\citeauthoryear{{Van de Sande}, {Decin}, {Lombaert}, {Khouri},
  {de Koter}, {Wyrowski}, {De Nutte}  \& {Homan}}{{Van de Sande}
  et~al.}{2018}]{VandeSande2018a}
{Van de Sande} M.,  {Decin} L.,  {Lombaert} R.,  {Khouri} T.,  {de Koter} A.,
  {Wyrowski} F.,  {De Nutte} R.,   {Homan} W.,  2018, \aap, 609, A63

\bibitem[\protect\citeauthoryear{{Van de Sande}, {Walsh}, {Mangan}  \&
  {Decin}}{{Van de Sande} et~al.}{2019}]{VandeSande2019b}
{Van de Sande} M.,  {Walsh} C.,  {Mangan} T.~P.,   {Decin} L.,  2019, \mn@doi
  [\mnras] {10.1093/mnras/stz2702}, \href
  {https://ui.adsabs.harvard.edu/abs/2019MNRAS.tmp.2325V} {p.~2325}

\bibitem[\protect\citeauthoryear{{Van de Sande}, {Walsh}  \& {Danilovich}}{{Van
  de Sande} et~al.}{2020}]{VandeSande2020}
{Van de Sande} M.,  {Walsh} C.,   {Danilovich} T.,  2020, \mn@doi [\mnras]
  {10.1093/mnras/staa1270}, \href
  {https://ui.adsabs.harvard.edu/abs/2020MNRAS.495.1650V} {495, 1650}

\bibitem[\protect\citeauthoryear{{Walsh}, {Millar}, {Nomura}, {Herbst},
  {Widicus Weaver}, {Aikawa}, {Laas}  \& {Vasyunin}}{{Walsh}
  et~al.}{2014}]{Walsh2014}
{Walsh} C.,  {Millar} T.~J.,  {Nomura} H.,  {Herbst} E.,  {Widicus Weaver} S.,
  {Aikawa} Y.,  {Laas} J.~C.,   {Vasyunin} A.~I.,  2014, \mn@doi [\aap]
  {10.1051/0004-6361/201322446}, \href
  {https://ui.adsabs.harvard.edu/abs/2014A&A...563A..33W} {563, A33}

\bibitem[\protect\citeauthoryear{{Whittet}}{{Whittet}}{2010}]{Whittet2010}
{Whittet} D.~C.~B.,  2010, \mn@doi [\apj] {10.1088/0004-637X/710/2/1009}, \href
  {https://ui.adsabs.harvard.edu/abs/2010ApJ...710.1009W} {710, 1009}

\bibitem[\protect\citeauthoryear{{Willner}, {Russell}, {Puetter}, {Soifer}  \&
  {Harvey}}{{Willner} et~al.}{1979}]{Willner1979}
{Willner} S.~P.,  {Russell} R.~W.,  {Puetter} R.~C.,  {Soifer} B.~T.,
  {Harvey} P.~M.,  1979, \mn@doi [\apjl] {10.1086/182931}, \href
  {https://ui.adsabs.harvard.edu/abs/1979ApJ...229L..65W} {229, L65}

\bibitem[\protect\citeauthoryear{{Winters}, {Fleischer}, {Le Bertre}  \&
  {Sedlmayr}}{{Winters} et~al.}{1997}]{Winters1997}
{Winters} J.~M.,  {Fleischer} A.~J.,  {Le Bertre} T.,   {Sedlmayr} E.,  1997,
  \aap, \href {https://ui.adsabs.harvard.edu/abs/1997A&A...326..305W} {326,
  305}

\bibitem[\protect\citeauthoryear{{Yang}, {Chen}  \& {He}}{{Yang}
  et~al.}{2004}]{Yang2004}
{Yang} X.,  {Chen} P.,   {He} J.,  2004, \mn@doi [\aap]
  {10.1051/0004-6361:20031673}, \href
  {https://ui.adsabs.harvard.edu/abs/2004A&A...414.1049Y} {414, 1049}

\bibitem[\protect\citeauthoryear{{Zhukovska} \& {Henning}}{{Zhukovska} \&
  {Henning}}{2013}]{Zhukovska2013}
{Zhukovska} S.,  {Henning} T.,  2013, \mn@doi [\aap]
  {10.1051/0004-6361/201321368}, \href
  {https://ui.adsabs.harvard.edu/abs/2013A&A...555A..99Z} {555, A99}

\makeatother
\end{thebibliography}




\appendix

\section{Details of the chemical kinetics model}			\label{app:model}

The chemical kinetics model is based on that of Papers I and II, and originates from the publicly available UMIST Database for Astrochemistry (UDfA) circumstellar envelope (CSE) model \citep{McElroy2013}\footnote{\url{http://udfa.ajmarkwick.net/index.php?mode=downloads}}.
The model uniquely describes both gas-phase and dust-gas chemistry: gas-phase species are able to accrete onto the dust (forming an ice mantle), react on the dust surface through the reactive Langmuir-Hinshelwood and the stick-and-hit Eley-Rideal mechanisms, and return to the gas phase via thermal desorption, photodesorption and sputtering (destroying the ice mantle).
The ice mantle is assumed to be physisorbed onto the dust, and not chemically integrated within the dust.
A complete description of the dust-gas chemistry and its reaction rates is given in Paper I.

The one-dimensional model describes a uniformly expanding spherically symmetric outflow, with constant mass-loss rate, $\dot{M}$, and gas outflow velocity, $v_\infty$.
The gas temperature profile is described by a power law,
\begin{equation}
    T_\mathrm{gas}(r) = T_* \left(\frac{r}{R_*}\right)^{-\epsilon},
\end{equation}
where $T_*$ and $R_*$ are the stellar temperature and radius.
Dust is assumed to be present at the start of the model, at $10^{15}$ cm $\approx$ 20 R$_*$.
Therefore, we assume that dust formation occurs in the inner wind and dust is injected into the intermediate wind with a certain grain size distribution (GSD) that does not change throughout the outflow. 
In this paper, the GSD is fixed to the canonical value of \cite[][MRN]{Mathis1977}.
The dust is assumed to have a fixed drift velocity, $v_\mathrm{drift}$, relative to the gas.
The dust temperature profile is described by 
\begin{equation}        \label{eq:tdust}
    T_\mathrm{dust}(r) = T_\mathrm{dust,*} \left( \frac{2r}{R_*} \right)^{-\frac{2}{4+s}},
\end{equation}
where $T_\mathrm{dust,*}$ and $s$ are free parameters.
Their values depend on the mass-loss rate, outflow velocity, and the composition of the dust. 
To derive them, we use the continuum radiative transfer code \textsc{MCMax} (see also Appendix \ref{app:tdustparams}).

\section{Influence of the new chemical network on the gas and ice phases}			\label{app:influencenew}

In this Appendix, we show the influence of including reactive desorption and grain surface photodissociation to the chemical reaction network.
We find that reactive desorption alone does not have a large influence on the total ice abundance throughout the outflow.
Together with grain-surface photodissociation, however, these two new reactions have a significant effect on the ice mantle.
Grain-surface photodissociation leads to more radical ice species, which can react via the Langmuir-Hinshelwood mechanism and subsequently desorb from the grain.
Additionally, the photodissociation products can be thermally desorbed if their binding energy is lower.
Since the ices are destroyed by photons, rather than photodesorbed in their entirety, the gas-phase abundance of the species will be lower, which leads to less reaccretion.

Sects. \ref{subapp:influencenew:orich} and \ref{subapp:influencenew:crich} describe the effects of the two new reactions on the ice component throughout an O-rich and a C-rich outflow, respectively.
In our notation, the prefix `G' (grain) denotes ice species.

\subsection{O-rich outflow}			\label{subapp:influencenew:orich}

\begin{figure}
 \includegraphics[width=1.\columnwidth]{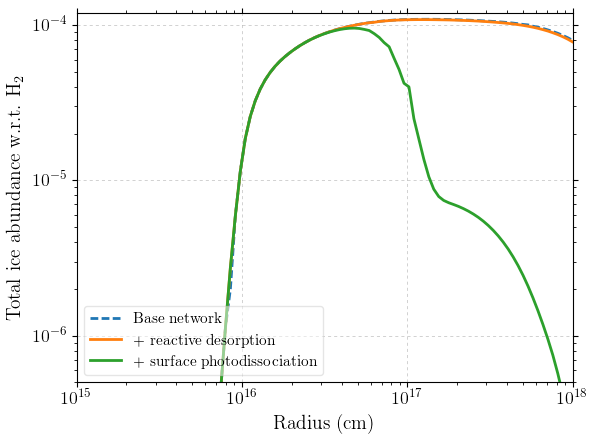}
 \caption{Total ice abundance relative to H$_2$ in an O-rich outflow with $\dot{M} = 10^{-5}$ M$_\odot$ yr$^{-1}$, $v_\infty = 15$ km s$^{-1}$ and $v_\mathrm{drift} = 5$ km s$^{-1}$.
 Blue dashed line: results from the updated chemical network. 
 Orange solid line: updated network + reactive desorption. 
 Green solid line: updated network + reactive desorption + grain surface photodissociation.
  }
 \label{fig:GP-total-orich}
\end{figure}

\begin{figure}
 \includegraphics[width=1.\columnwidth]{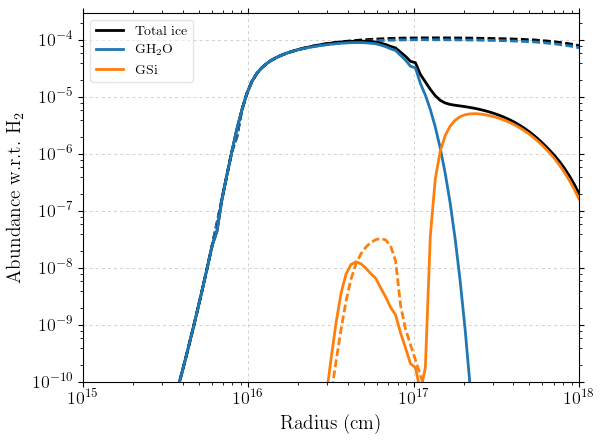}
 \caption{Fractional abundance of key ice species in an O-rich outflow with $\dot{M} = 10^{-5}$ M$_\odot$ yr$^{-1}$, $v_\infty = 15$ km s$^{-1}$ and $v_\mathrm{drift} = 5$ km s$^{-1}$.
 Dashed lines: results from the updated chemical network.
 Solid lines: results from the updated network + reactive desorption + grain surface photodissociation.
 The total fractional ice abundance is plotted in black.
  }
 \label{fig:GP-evol-orich}
\end{figure}

Fig. \ref{fig:GP-total-orich} shows the total ice abundance relative to H$_2$ in an O-rich outflow, characterised by $\dot{M} = 10^{-5}$ M$_\odot$ yr$^{-1}$, $v_\infty = 15$ km s$^{-1}$ and $v_\mathrm{drift} = 5$ km s$^{-1}$ for different chemical networks.
We show the results using the base chemical network, which includes all reactions described in Papers I and II, the base network including reactive desorption, and the base network including reactive desorption and grain-surface photodissociation.
Fig. \ref{fig:GP-evol-orich} shows the fractional abundances of the main ice components for the base network and the network including both reactions.

Including reactive desorption has a very small effect on the total ice abundance in the outer region of the outflow, where the ice abundance of radicals increases due to accretion only in this network.
When using the base network and the network including reactive desorption, the main ice component is GH$_2$O throughout the outflow.
When including grain-surface photodissociation, GH$_2$O is broken up, providing GOH radicals.
Two of these radicals can react, forming GH$_2$O$_2$, which subsequently desorbs upon reaction.
This removes water both from the ice and the gas phase, hindering its reaccreation.
The gas-phase abundance of H$_2$O$_2$ also increases. 
For this dense outflow, we find that its peak abundance around $10^{17}$ cm (i.e., before the onset of photodissociation) increases from $\sim 10^{-10}$ to $\sim 10^{-5}$ relative to H$_2$. 
The peak abundance decreases to $\sim 10^{-7}$ relative to H$_2$ for lower density outflows. 

After the removal of GH$_2$O, the main ice component depends on the outflow density. 
In high-density outflows, the main ice component is GSi.
This is the result of grain surface photodissociation of GSiO, and to a lesser extent GSiS. 
While GO is very reactive, GSi remains on the grain.
For outflows with $\dot{M} < 10^{-5}$ M$_\odot$ yr$^{-1}$, the main component is GSiH$_4$ due to a larger GH abundance thanks to the onset of photon-processes closer to the star.
This leads to an increase in its gas-phase abundance. 
The peak gas-phase abundance of SiH$_4$ ranges with increasing outflow density from $\sim 10^{-12}$ to $\sim 10^{-8}$ relative to H$_2$.

\subsection{C-rich outflow}			\label{subapp:influencenew:crich}

\begin{figure}
 \includegraphics[width=1.\columnwidth]{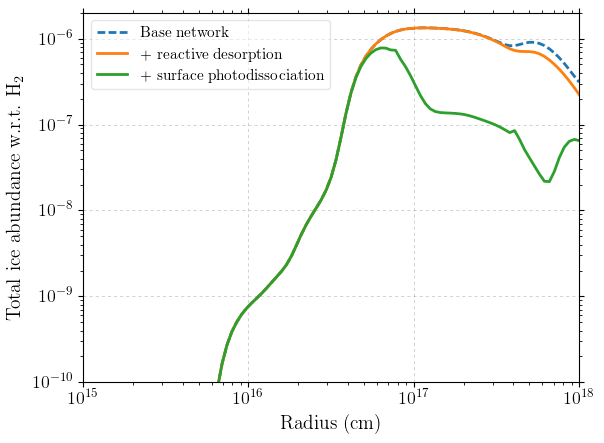}
 \caption{Total ice abundance relative to H$_2$ in an C-rich outflow with $\dot{M} = 10^{-5}$ M$_\odot$ yr$^{-1}$, $v_\infty = 15$ km s$^{-1}$ and $v_\mathrm{drift} = 5$ km s$^{-1}$, using the parents species used in Papers I and II.
 Blue dashed line: results from the updated chemical network. 
 Orange solid line: updated network + reactive desorption. 
 Green solid line: updated network + reactive desorption + grain surface photodissociation.
  }
 \label{fig:GP-total-crich}
\end{figure}

\begin{figure}
 \includegraphics[width=1.\columnwidth]{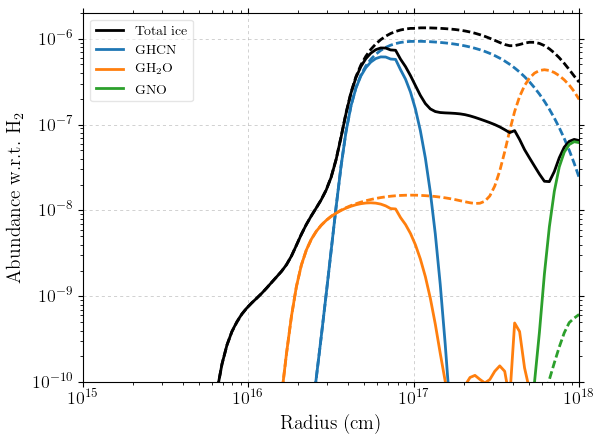}
 \caption{Fractional abundances of key ice species in an C-rich outflow with $\dot{M} = 10^{-5}$ M$_\odot$ yr$^{-1}$, $v_\infty = 15$ km s$^{-1}$, and $v_\mathrm{drift} = 5$ km s$^{-1}$, using the parents species used in Papers I and II.
 Dashed lines: results from the updated chemical network.
 Solid lines: results from the updated network + reactive desorption + grain surface photodissociation.
 The total fractional ice abundance is plotted in black.
  }
 \label{fig:GP-evol-crich}
\end{figure}

Similar to the O-rich outflow, Fig. \ref{fig:GP-total-crich} shows the total ice abundance relative to H$_2$ in a C-rich outflow, characterised by $\dot{M} = 10^{-5}$ M$_\odot$ yr$^{-1}$, $v_\infty = 15$ km s$^{-1}$ and $v_\mathrm{drift} = 5$ km s$^{-1}$ for different chemical networks, using the parents species used in Papers I and II.
Fig. \ref{fig:GP-evol-crich} shows the fractional abundances of its main ice components for the base network and the network including both reactions.

Including reactive desorption again only has a small effect on the total ice abundance in the outer region of the outflow.
Using the base model and the model including reactive desorption, the main ice components are GHCN and GH$_2$O.
The bump in the ice abundance in the outer outflow is due to GH$_2$O.
It is not present when using the chemical network of Papers I and II and is formed thanks to the more complex surface chemical network.
When including grain surface photodissociation, GHCN and GH$_2$O are photodissociated, yielding GOH + GOH (which reactively desorbs, see Sect. \ref{subapp:influencenew:orich}) and GH + GCN.
The peak in GNO at the end of the outflow is linked to the surface photodissociation of GHCN, via the reaction sequence
\begin{equation*}
\begin{split}
   & \mathrm{GO} + \mathrm{GCN} \rightarrow \mathrm{GOCN},\\
   & \mathrm{GH} + \mathrm{GOCN} \rightarrow \mathrm{GHOCN},\\
   & \mathrm{GHOCN} + h\nu \rightarrow \mathrm{GOH} + \mathrm{GNH},\\
   & \mathrm{GO} + \mathrm{GNH} \rightarrow \mathrm{GHNO,}\\
   & \mathrm{GO} + \mathrm{GHNO} \rightarrow \mathrm{GNO} + \mathrm{GOH}.
\end{split}
\end{equation*}

\section{Parameters of the observational grids of outflows}			\label{app:tdustparams}

Tables \ref{table:paramscrich} and \ref{table:paramsorich} list the observational set of C-rich and O-rich outflows, respectively, with the observed mass-loss rates and outflow velocities and the assumed drift velocities.
As in Papers I and II, the parameters $T_\mathrm{dust,*}$ and $s$ of the dust temperature power-law are obtained by fitting Eq. \ref{eq:tdust} to the dust temperature profiles calculated using the continuum radiative transfer code \textsc{MCMax} \citep{Min2009}.
This was done by identifying the minimum value of the $\chi^2$ test when varying  $T_\mathrm{dust,*}$ in intervals of 50 K and $s$ in intervals of 0.1. 
The minimum $\chi^2$ values of the fitting routine are also included in Tables \ref{table:outputobscrich} and \ref{table:outputobsorich}.
The optical constants used in the radiative transfer modelling for the O-rich outflows are those of \citet{Mutschke1998} for melilite and those of \citet{Jager2003} for silicate with Fe.
For the C-rich outflows, we used the optical constants of \citet{Preibisch1993} for amorphous carbon.

\section{Influence on the gas-phase composition}		\label{app:gas}

The influence on the gas-phase composition is the main topic of Paper I, where we found that including dust-gas chemistry can significantly influence the gas phase throughout the outflow.
The depletion of parent species depends on the outflow density, the drift velocity, the dust temperature as well as the GSD (Paper II).

Consistent with the results of Papers I and II, the depletion of parent species is smaller than one order of magnitude for both observational sets of outflows. 
This is in part due to the low drift velocities assumed for the highest mass-loss rates, the assumption of the canonical MRN GSD as well as the warmer 50/50 mixture of melilite and silicate with Fe dust used in the O-rich calculations.

Daughter species can be affected by dust-gas chemistry in two ways. 
They can be depleted onto the dust, which occurs more efficiently for species with a large binding energy, or experience an increase in abundance because of their formation on the dust, followed by thermal desorption, photodesorption or sputtering.
Because of the more complex grain-surface chemical network, the different species can form on the grain surface, leading to a different gas-phase return compared to Paper I.

\begin{figure}
 \includegraphics[width=1.\columnwidth]{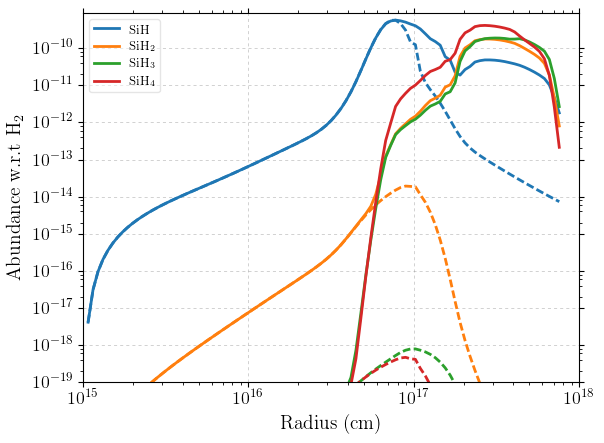}
 \caption{Abundance profiles of daughter species produced by grain-surface chemistry in a C-rich outflow with $\dot{M} = 10^{-5}$ M$_\odot$ yr$^{-1}$, $v_\infty = 15$ km s$^{-1}$ and $v_\mathrm{drift} = 5$ km s$^{-1}$, using the parent species of Paper I.
 Solid lines: results obtained when including the full chemical network.
 Dashed lines: results obtained without including dust-gas chemistry.
  }
 \label{fig:gp-crich}
\end{figure}

\begin{figure}
 \includegraphics[width=1.\columnwidth]{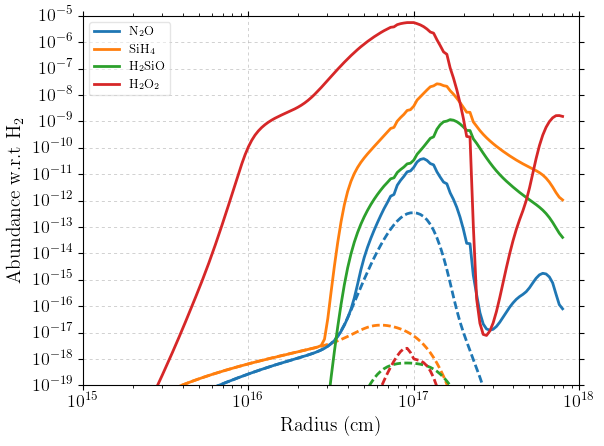}
 \caption{Abundance profiles of daughter species produced by grain-surface chemistry in a O-rich outflow with $\dot{M} = 10^{-5}$ M$_\odot$ yr$^{-1}$, $v_\infty = 15$ km s$^{-1}$ and $v_\mathrm{drift} = 5$ km s$^{-1}$.
 Solid lines: results obtained when including the full chemical network.
 Dashed lines: results obtained without including dust-gas chemistry.
 }
 \label{fig:gp-orich}
\end{figure}

Figs. \ref{fig:gp-crich} and \ref{fig:gp-orich} show gas-phase abundance profiles of daughter species produced on grain surfaces for a C-rich and O-rich outflow, respectively. 
Hydrates are efficiently formed on the grain surface thanks to the high mobility of H and are subsequently thermally desorbed to the gas phase. 
The largest influence is seen for SiH$_n$ ($n=1,2,3,4$) in both C-rich and O-rich outflows.
Reactions with other mobile atoms, such as O, also lead to gas-phase increases (e.g., NO$_2$ in O-rich outflows).
As mentioned in Appendix \ref{app:influencenew}, H$_2$O$_2$ is released into the gas phase via reactive desorption in O-rich outflows.

Thanks to the complex grain-surface chemical network, various complex ices can be formed on the grain and are released to the gas phase via reactive desorption.
Although this increases the gas-phase abundances of simple and complex organic molecules, with surface reactions being the only formation route for some (e.g., CH$_3$COOH), their peak gas-phase abundances and column densities remain low.
Not including photoprocessing increases their gas-phase abundances, but does not lead to detectable peak gas-phase abundances or column densities.
The influence on the gas-phase abundance of C-bearing species is therefore limited due to the extended chemical network, as it distributes the total carbon budget over many species.

\section{Refractory organic output of O-rich outflows}			\label{app:outputorich}

\begin{figure}
 \includegraphics[width=1.\columnwidth]{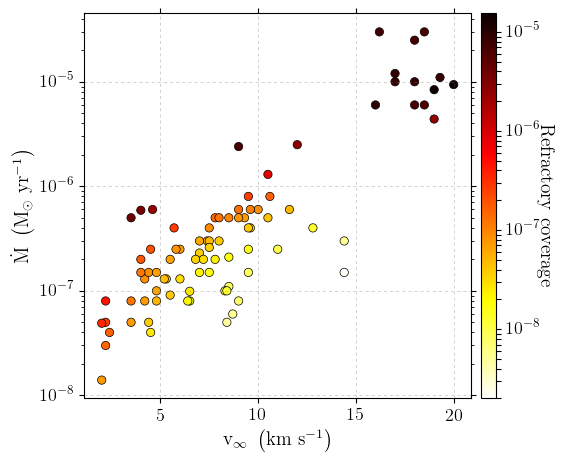}
 \caption{Refractory organic coverage of the dust grains in the outflows of the O-rich observational grid.
 The coverage corresponds to the number of refractory organic monolayers where the outflow merges with the ISM, which is taken to be the radius where $n(\mathrm{H_2}) \leq 10$ cm$^{-3}$.
   }
 \label{fig:outputobsorich}
\end{figure}

Fig. \ref{fig:outputobsorich} shows the refractory organic coverage of the dust in the O-rich observational grid.
The average refractory organic output to the ISM is listed in Table \ref{table:outputobsorich}.
The refractory organic coverage again increases with increasing outflow density.
Dust grains in O-rich outflows have a refractory organic coverage about four orders of magnitude smaller compared to C-rich outflows.

\begin{figure*}
 \includegraphics[width=1.0\textwidth]{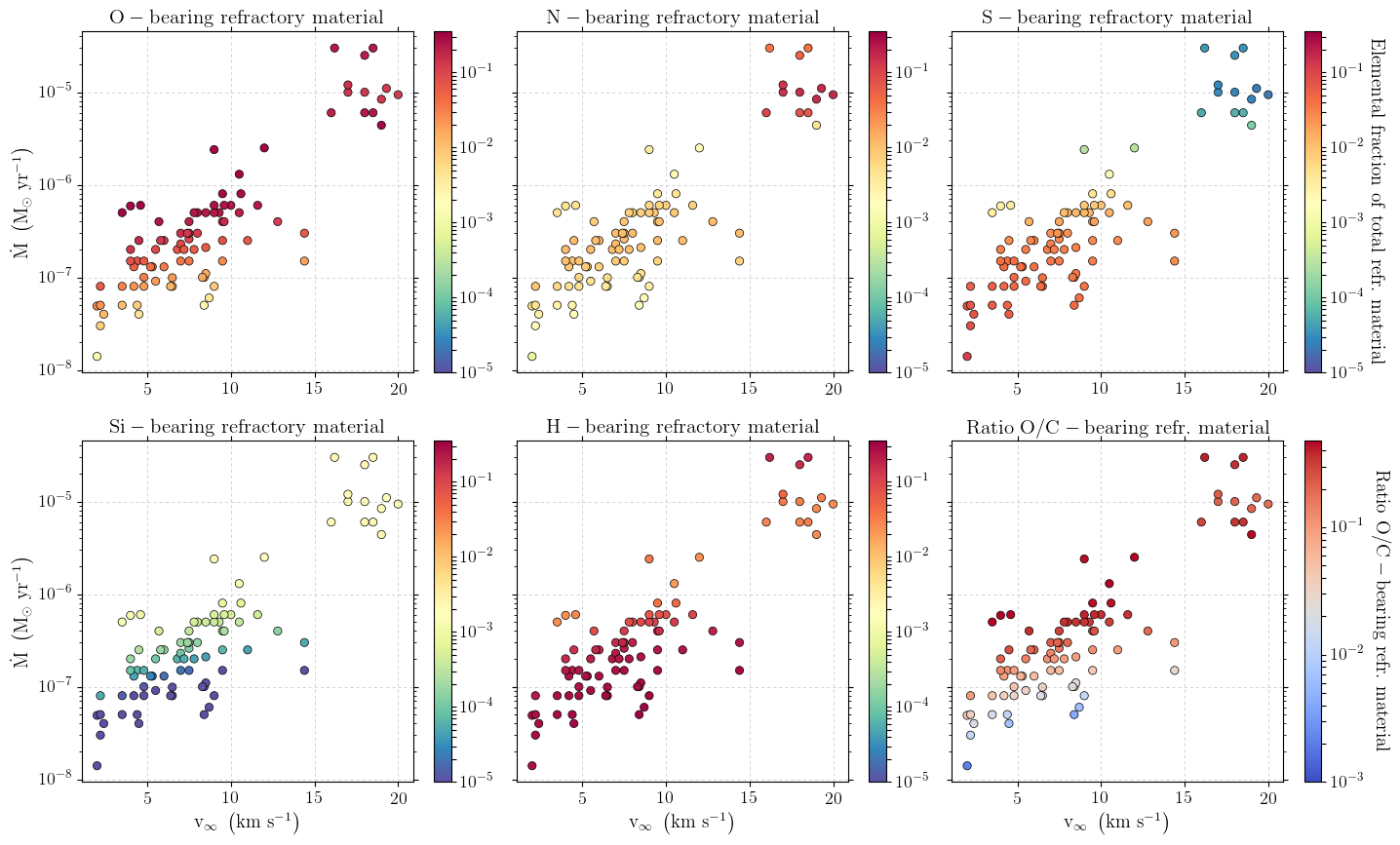}
 \caption{First five panels: elemental fractional composition of the total refractory organic output (O-, N-, S-, Si-, and H-bearing refractory organic material) for the outflows of the O-rich observational dataset.
 Lower right panel: ratio of O-bearing/C-bearing refractory organic material of the total refractory organic output.
   }
 \label{fig:outputobsorich-allfracs}
\end{figure*}

Fig. \ref{fig:outputobsorich-allfracs} shows the fractional composition of the refractory organic material.
The same processes are at play as in the C-rich outflows: accretion of gas-phase daughter species followed by the onset of photon-reactions, where the onset depends on the outflow density.
We find that the same trends with increasing outflow density are present for the O-rich and the C-rich outflows (Sect. \ref{subsect:results:crich} and Fig. \ref{subsect:discussion:complexity}), except for the H-bearing component.

In the O-rich outflows, the fraction of H-bearing species decreases with increasing outflow density.
C-bearing daughter species are formed in the gas phase, though with much smaller abundances than in C-rich outflows.
These are accreted more efficiently on the colder dust and photoprocessed closer to the star in lower density outflows, leading to larger component C-chains, part of which are H-bearing, in the ice and refractory organic material.
To a lesser extent, the decreasing fraction of H-bearing species follows the increased fraction of O-bearing species.
Because of the larger H$_2$O abundance, the amount of O-bearing radicals on the grain is larger, leading to a larger refractory organic component that increases with outflow density. 
The main O-bearing components do not contain any H (GC$_n$O, $n = 2,3$).

The reversal in the trends for outflows with $\dot{M} > 10^{-5}$ M$_\odot$ yr$^{-1}$ is due to the later onset of CO and HCN photodissociation, which shifts to the outer edge of the wind within these outflows.
The larger CO envelope leads to a larger GCO abundance, which is also less likely to be photodissociated into GC and GO.
A larger GCO abundance leads to a larger GC$_n$O abundance via reactions with GC, increasing the fraction of O-bearing species in the refractory organic mantle.
The larger GHCN abundance is less likely to be photodissociated into GH and GCN.
The latter is the main reactant involved in the formation of the N-bearing component.

\newpage

\setcounter{table}{0}
\renewcommand{\thetable}{C.\arabic{table}}

\begin{table*}
	\caption{Observational set of C-rich outflows with their observed mass-loss rates, $\dot{M}$ [M$_\odot$ yr$^{-1}$], and expansion velocities, $v_\infty$ [km s$^{-1}$], from \citet{Schoier2001} and \citet{Danilovich2015} and the assumed drift velocities, $v_\mathrm{drift}$ [km s$^{-1}$].
	The values of the free parameters of the dust temperature profile $T_\mathrm{dust,*}$ and $s$, together with the chi square value of each fit, are also listed.
	}
	\centering
	\label{table:paramscrich}
	\begin{tabular}{lcccccclcccccc} 
	\hline
	Name & $\dot{M}$ & $v_\infty$ & $v_\mathrm{drift}$ & $T_\mathrm{dust,*}$ & $s$ & $\chi^2$ & Name & $\dot{M}$ & $v_\infty$ & $v_\mathrm{drift}$ & $T_\mathrm{dust,*}$ & $s$ & $\chi^2$ \\
	\hline
    \noalign{\smallskip}
HD 124268 & 1e-07 & 11.0 & 15.0 & 2050 & 0.8 & 0.35 & SS Vir & 2e-07 & 12.5 & 10.0 & 2000 & 0.8 & 0.23\\
CZ Hya & 9e-07 & 12.0 & 10.0 & 1850 & 0.8 & 1.87 & Y Pav & 1.6e-07 & 8.0 & 10.0 & 2000 & 0.8 & 3.3\\
V460 Cyg & 1.8e-07 & 10.0 & 10.0 & 2000 & 0.8 & 1.82 & U Hya & 8.9e-08 & 6.5 & 15.0 & 2000 & 0.8 & 1.47\\
X Vel & 1.8e-07 & 10.0 & 10.0 & 2000 & 0.8 & 1.82 & TW Oph & 5e-08 & 7.5 & 15.0 & 2050 & 0.8 & 1.54\\
BL Ori & 1.1e-07 & 9.0 & 10.0 & 2000 & 0.8 & 3.74 & XZ Vel & 6e-07 & 14.0 & 10.0 & 1900 & 0.8 & 0.47\\
UU Aur & 1.7e-07 & 12.0 & 10.0 & 2000 & 0.8 & 1.14 & TU Gem & 2.5e-07 & 11.5 & 10.0 & 1950 & 0.8 & 2.54\\
Y Hya & 1.9e-07 & 9.0 & 10.0 & 1950 & 0.8 & 4.5 & CL Mon & 2.2e-06 & 25.0 & 10.0 & 1800 & 0.8 & 3.28\\
GY Cam & 3.7e-06 & 20.0 & 10.0 & 1750 & 0.8 & 2.38 & V1426 Cyg & 1e-05 & 14.0 & 5.0 & 1800 & 0.6 & 5.37\\
RV Cyg & 4.5e-07 & 13.5 & 10.0 & 1900 & 0.8 & 4.92 & CCS 2792 & 6.5e-07 & 17.0 & 10.0 & 1900 & 0.8 & 0.7\\
HV Cas & 9e-07 & 18.5 & 10.0 & 1900 & 0.8 & 3.06 & II Lup & 1.7e-05 & 21.5 & 5.0 & 1800 & 0.6 & 4.85\\
TT Tau & 7e-08 & 5.0 & 15.0 & 2000 & 0.8 & 1.25 & V1968 Cyg & 7.5e-06 & 20.0 & 10.0 & 1750 & 0.7 & 2.57\\
V384 Per & 2.6e-06 & 15.0 & 10.0 & 1750 & 0.8 & 2.16 & RW Lmi & 6e-06 & 17.0 & 10.0 & 1750 & 0.7 & 3.13\\
U Cyg & 9e-07 & 13.0 & 10.0 & 1850 & 0.8 & 0.85 & Y Cvn & 1.5e-07 & 8.5 & 10.0 & 2000 & 0.8 & 0.99\\
V Aql & 1.3e-07 & 11.0 & 10.0 & 2000 & 0.8 & 3.66 & RV Aqr & 2.3e-06 & 15.0 & 10.0 & 1850 & 0.7 & 1.56\\
T Dra & 1.2e-06 & 13.5 & 10.0 & 1800 & 0.8 & 3.13 & CW Leo & 1.5e-05 & 14.5 & 5.0 & 1800 & 0.6 & 9.14\\
W Cma & 3e-07 & 10.5 & 10.0 & 1950 & 0.8 & 0.38 & X Cnc & 1e-07 & 7.0 & 15.0 & 2000 & 0.8 & 0.87\\
V688 Mon & 6.1e-06 & 13.5 & 10.0 & 1750 & 0.7 & 1.55 & W Pic & 3e-07 & 16.0 & 10.0 & 2000 & 0.8 & 1.81\\
V821 Her & 3e-06 & 13.5 & 10.0 & 1800 & 0.7 & 1.88 & V701 Cas & 4.5e-06 & 11.5 & 10.0 & 1750 & 0.7 & 2.37\\
IRC +60041 & 3e-06 & 28.0 & 10.0 & 1800 & 0.8 & 1.24 & T Ind & 9e-08 & 6.0 & 15.0 & 2000 & 0.8 & 0.48\\
RT Cap & 1e-07 & 8.0 & 15.0 & 2000 & 0.8 & 3.0 & AQ Sgr & 2.5e-07 & 10.0 & 10.0 & 1950 & 0.8 & 0.23\\
S Aur & 2.2e-06 & 25.5 & 10.0 & 1800 & 0.8 & 2.56 & R Lep & 8.7e-06 & 18.0 & 10.0 & 1750 & 0.7 & 2.14\\
WZ Cas & 6.5e-09 & 2.5 & 15.0 & 2100 & 0.8 & 0.46 & V466 Per & 1.3e-07 & 9.0 & 10.0 & 2000 & 0.8 & 0.87\\
R For & 1.3e-06 & 16.5 & 10.0 & 1850 & 0.8 & 5.08 & W Ori & 7e-08 & 11.0 & 15.0 & 2050 & 0.8 & 1.81\\
Y Tau & 4e-07 & 11.0 & 10.0 & 1900 & 0.8 & 2.79 & SZ Car & 2e-06 & 14.0 & 10.0 & 1850 & 0.7 & 1.95\\
X TrA & 1.9e-07 & 6.5 & 10.0 & 1950 & 0.8 & 1.66 & V1259 Ori & 8.8e-06 & 16.0 & 10.0 & 1750 & 0.7 & 5.01\\
RY Mon & 3.5e-07 & 11.0 & 10.0 & 1950 & 0.8 & 3.54 & ST Cam & 9e-08 & 9.0 & 15.0 & 2050 & 0.8 & 1.7\\
AI Vol & 4.9e-06 & 12.0 & 10.0 & 1750 & 0.7 & 1.82 & Z Psc & 2.5e-08 & 3.5 & 15.0 & 2050 & 0.8 & 0.87\\
S Cep & 1.5e-06 & 22.0 & 10.0 & 1850 & 0.8 & 1.04 & V Cyg & 1.2e-06 & 11.5 & 10.0 & 1800 & 0.8 & 1.26\\
R Vol & 1.8e-06 & 18.0 & 10.0 & 1800 & 0.8 & 1.1 & HD 121658 & 1e-07 & 6.5 & 15.0 & 2000 & 0.8 & 0.38\\
LP And & 1.5e-05 & 14.0 & 5.0 & 1800 & 0.6 & 10.05 & TW Hor & 9e-08 & 5.5 & 15.0 & 2000 & 0.8 & 0.36\\
V CrB & 3.3e-07 & 7.5 & 10.0 & 1900 & 0.8 & 1.11 & VX And & 4e-08 & 11.5 & 15.0 & 2100 & 0.8 & 0.47\\
PQ Cep & 1.4e-06 & 19.5 & 10.0 & 1850 & 0.8 & 1.95 & RY Dra & 3e-07 & 10.0 & 10.0 & 1950 & 0.8 & 1.59\\
T Lyr & 7e-08 & 11.5 & 15.0 & 2050 & 0.8 & 3.26 & NP Pup & 6.5e-08 & 9.5 & 15.0 & 2050 & 0.8 & 1.23\\
VY Uma & 7e-08 & 6.0 & 15.0 & 2000 & 0.8 & 4.35 & V1942 Sgr & 1.6e-07 & 10.0 & 10.0 & 2000 & 0.8 & 0.23\\
UX Dra & 1.6e-07 & 4.0 & 10.0 & 1900 & 0.8 & 0.5 & & & & & & \\
	\hline
	\end{tabular}
\end{table*}

\begin{table*}
	\caption{Observational set of O-rich outflows with their observed mass-loss rates, $\dot{M}$ [M$_\odot$ yr$^{-1}$], and expansion velocities, $v_\infty$ [km s$^{-1}$], from \citet{Olofsson2002}, \citet{GonzalezDelgado2003}, and \citet{Danilovich2015} and the assumed drift velocities, $v_\mathrm{drift}$ [km s$^{-1}$].
	The values of the free parameters of the dust temperature profile $T_\mathrm{dust,*}$ and $s$, together with the chi square value of each fit, are also listed.
	}
	\centering
	\label{table:paramsorich}
	\begin{tabular}{lcccccclcccccc} 
	\hline
	Name & $\dot{M}$ & $v_\infty$ & $v_\mathrm{drift}$ & $T_\mathrm{dust,*}$ & $s$ & $\chi^2$ & Name & $\dot{M}$ & $v_\infty$ & $v_\mathrm{drift}$ & $T_\mathrm{dust,*}$ & $s$ & $\chi^2$ \\
	\hline
    \noalign{\smallskip}
RY CrB & 4e-07 & 5.7 & 10.0 & 950 & 1.3 & 35.99 & L$_2$ Pup & 1.4e-08 & 2.0 & 15.0 & 950 & 1.3 & 33.19\\
FZ Hya & 2e-07 & 7.8 & 10.0 & 950 & 1.3 & 34.69 & AZ UMa & 2.5e-07 & 4.5 & 10.0 & 950 & 1.3 & 35.41\\
IRC+50137 & 1e-05 & 17.0 & 5.0 & 950 & 1.3 & 40.38 & IK Tau & 3e-05 & 18.5 & 5.0 & 950 & 1.3 & 55.88\\
RX Lac & 8e-08 & 2.2 & 15.0 & 950 & 1.3 & 34.73 & UU Dra & 5e-07 & 8.0 & 10.0 & 950 & 1.3 & 35.21\\
IRC-30398 & 6e-06 & 16.0 & 10.0 & 950 & 1.3 & 39.71 & WX Psc & 1.1e-05 & 19.3 & 5.0 & 950 & 1.3 & 41.92\\
TZ Aql & 1e-07 & 4.8 & 15.0 & 950 & 1.3 & 33.37 & UX Sgr & 1.5e-07 & 9.5 & 10.0 & 950 & 1.3 & 34.63\\
AK Hya & 1e-07 & 4.8 & 15.0 & 950 & 1.3 & 33.37 & T Mic & 8e-08 & 4.8 & 15.0 & 950 & 1.3 & 34.6\\
V1943 Sgr & 9.9e-08 & 6.5 & 15.0 & 950 & 1.3 & 35.07 & FK Hya & 6e-08 & 8.7 & 15.0 & 950 & 1.3 & 33.29\\
RW Vir & 1.5e-07 & 7.0 & 10.0 & 950 & 1.3 & 33.5 & SW Vir & 4e-07 & 7.5 & 10.0 & 950 & 1.3 & 35.53\\
BI Car & 3e-08 & 2.2 & 15.0 & 950 & 1.3 & 34.1 & S CMi & 4.9e-08 & 2.0 & 15.0 & 950 & 1.3 & 33.71\\
R Dor & 1.3e-07 & 6.0 & 10.0 & 950 & 1.3 & 33.6 & TX Cam & 6e-06 & 18.5 & 10.0 & 950 & 1.3 & 39.89\\
TW Peg & 2.5e-07 & 9.5 & 10.0 & 950 & 1.3 & 34.12 & R LMi & 2.6e-07 & 7.5 & 10.0 & 950 & 1.3 & 35.29\\
S Pav & 8e-08 & 9.0 & 15.0 & 950 & 1.3 & 32.89 & CZ Ser & 8e-07 & 9.5 & 10.0 & 950 & 1.3 & 37.32\\
CS Dra & 6e-07 & 11.6 & 10.0 & 950 & 1.3 & 34.78 & RX Lep & 5e-08 & 3.5 & 15.0 & 950 & 1.3 & 33.98\\
V744 Cen & 1.3e-07 & 5.3 & 10.0 & 950 & 1.3 & 33.91 & V352 Ori & 5e-08 & 8.4 & 15.0 & 950 & 1.3 & 33.74\\
SY Lyr & 6e-07 & 4.6 & 10.0 & 950 & 1.3 & 38.71 & V PsA & 3e-07 & 14.4 & 10.0 & 950 & 1.3 & 33.37\\
UY Cet & 2.5e-07 & 6.0 & 10.0 & 950 & 1.3 & 35.77 & SV Peg & 3e-07 & 7.5 & 10.0 & 950 & 1.3 & 35.25\\
RX Boo & 5e-07 & 9.3 & 10.0 & 950 & 1.3 & 34.98 & R Hor & 5.9e-07 & 4.0 & 10.0 & 950 & 1.3 & 37.94\\
CE And & 5e-07 & 10.5 & 10.0 & 950 & 1.3 & 36.73 & AB Aqr & 1.3e-07 & 4.2 & 10.0 & 950 & 1.3 & 36.22\\
V1111 Oph & 1.2e-05 & 17.0 & 5.0 & 950 & 1.3 & 42.79 & RW Lep & 5e-08 & 4.4 & 15.0 & 950 & 1.3 & 33.6\\
W Hya & 8e-08 & 6.5 & 15.0 & 950 & 1.3 & 33.78 & RV Cam & 2.5e-07 & 5.8 & 10.0 & 950 & 1.3 & 35.6\\
Y Scl & 1.3e-07 & 5.2 & 10.0 & 950 & 1.3 & 34.11 & AFGL 292 & 2.1e-07 & 8.5 & 10.0 & 950 & 1.3 & 34.46\\
SU Vel & 2e-07 & 5.5 & 10.0 & 950 & 1.3 & 34.73 & SZ Dra & 6e-07 & 9.6 & 10.0 & 950 & 1.3 & 35.21\\
V Tel & 2e-07 & 6.8 & 10.0 & 950 & 1.3 & 35.76 & R Leo & 1.1e-07 & 8.5 & 10.0 & 950 & 1.3 & 33.71\\
Y Tel & 5e-07 & 3.5 & 10.0 & 1000 & 1.2 & 39.69 & W Cyg & 1e-07 & 8.3 & 15.0 & 950 & 1.3 & 33.39\\
BX Cam & 4.4e-06 & 19.0 & 10.0 & 950 & 1.3 & 38.15 & CW Cnc & 5e-07 & 8.5 & 10.0 & 950 & 1.3 & 34.17\\
SU Sgr & 4e-07 & 9.5 & 10.0 & 950 & 1.3 & 35.3 & IRC+10365 & 3e-05 & 16.2 & 5.0 & 950 & 1.3 & 53.92\\
GX Mon & 8.4e-06 & 19.0 & 10.0 & 950 & 1.3 & 38.66 & EY Hya & 2.5e-07 & 11.0 & 10.0 & 950 & 1.3 & 34.09\\
AF Cyg & 8e-08 & 3.5 & 15.0 & 950 & 1.3 & 34.54 & T Ari & 4e-08 & 2.4 & 15.0 & 950 & 1.3 & 34.6\\
T Cep & 9.1e-08 & 5.5 & 15.0 & 950 & 1.3 & 34.89 & BC And & 2e-07 & 4.0 & 10.0 & 950 & 1.3 & 36.32\\
SV Aqr & 3e-07 & 8.0 & 10.0 & 950 & 1.3 & 34.35 & S CrB & 2.3e-07 & 7.0 & 10.0 & 950 & 1.3 & 35.92\\
KU And & 9.4e-06 & 20.0 & 10.0 & 950 & 1.3 & 42.04 & UX And & 4e-07 & 12.8 & 10.0 & 950 & 1.3 & 35.75\\
RS And & 1.5e-07 & 4.4 & 10.0 & 950 & 1.3 & 36.61 & S Dra & 4e-07 & 9.6 & 10.0 & 950 & 1.3 & 35.77\\
g Her & 1e-07 & 8.4 & 15.0 & 950 & 1.3 & 33.15 & IRC-10529 & 2.5e-06 & 12.0 & 10.0 & 950 & 1.3 & 37.99\\
Y UMa & 1.5e-07 & 4.8 & 10.0 & 950 & 1.3 & 35.75 & BK Vir & 1.5e-07 & 4.0 & 10.0 & 950 & 1.3 & 34.35\\
R Hya & 3e-07 & 7.0 & 10.0 & 950 & 1.3 & 35.3 & RR Aql & 2.4e-06 & 9.0 & 10.0 & 950 & 1.3 & 38.29\\
SS Cep & 6e-07 & 10.0 & 10.0 & 950 & 1.3 & 34.15 & V584 Aql & 5e-08 & 2.2 & 15.0 & 950 & 1.3 & 34.09\\
R Cas & 1.3e-06 & 10.5 & 10.0 & 950 & 1.3 & 37.3 & TU Lyr & 3e-07 & 7.4 & 10.0 & 950 & 1.3 & 34.97\\
EX Ori & 8e-08 & 4.2 & 15.0 & 950 & 1.3 & 34.09 & Theta Aps & 4e-08 & 4.5 & 15.0 & 950 & 1.3 & 32.89\\
R Crt & 8e-07 & 10.6 & 10.0 & 950 & 1.3 & 35.98 & IRC +40004 & 6e-06 & 18.0 & 10.0 & 950 & 1.3 & 42.63\\
OT Pup & 5e-07 & 9.0 & 10.0 & 950 & 1.3 & 35.41 & RT Vir & 5e-07 & 7.8 & 10.0 & 950 & 1.3 & 36.06\\
AH Dra & 8e-08 & 6.4 & 15.0 & 950 & 1.3 & 33.43 & TY Dra & 6e-07 & 9.0 & 10.0 & 950 & 1.3 & 37.08\\
$\tau^4$ Ser & 1.5e-07 & 14.4 & 10.0 & 950 & 1.3 & 33.72 & U Del & 1.5e-07 & 7.5 & 10.0 & 950 & 1.3 & 33.36\\
V1300 Aql & 1e-05 & 18.0 & 5.0 & 950 & 1.3 & 41.08 & U Men & 2e-07 & 7.2 & 10.0 & 950 & 1.3 & 36.2\\
NV Aur & 2.5e-05 & 18.0 & 5.0 & 950 & 1.3 & 52.82 & & & & & &\\
	\hline
	\end{tabular}
\end{table*}



\bsp	
\label{lastpage}
\end{document}